\documentclass[twocolumn]{autart}    % Enable this line and disable the
\usepackage{amsmath}
\usepackage{mathrsfs}
\usepackage{amssymb}
\usepackage{amsfonts}
\usepackage{mathtools}
\usepackage{bm}
\usepackage[normalem]{ulem}  % SUPER!!! AVOIDED UNDERLYING WHEN \emph
\usepackage{fancyhdr}
\usepackage{hyperref}
\usepackage{framed}
\usepackage{float}
\usepackage[pdftex,final]{graphicx}
\usepackage{tikz}
\usepackage{comment}
\usepackage{sidecap}
\usepackage{xcolor}
\usepackage{yfonts}
\usepackage{enumitem}   
\usepackage{pgfplots}
\usetikzlibrary{calc,arrows}
\usetikzlibrary{patterns}

\newtheorem{corollary}{Corollary}
\newtheorem{lemma}{Lemma}
\newtheorem{assumption}{Assumption}
\newtheorem{remark}{Remark}
\newtheorem{dfn}{Definition}

\newcommand{\mytilde}{\raise.17ex\hbox{$\scriptstyle\mathtt{\sim}$}}

\newcommand{\I}{\ensuremath{\mathcal{I}}}
\newcommand{\bfI}{\ensuremath{\boldsymbol{\mathcal{I}}}}
\renewcommand{\S}{\ensuremath{\mathcal{S}}}
\renewcommand{\P}{\ensuremath{\mathcal{P}}}
\newcommand{\R}{\ensuremath{\mathcal{R}}}

\newcommand{\N}{\ensuremath{\mathcal{N}}}

\newcommand{\G}{\ensuremath{\mathcal{G}}}

\newcommand{\M}{\ensuremath{\mathcal{M}}}

\newcommand{\U}{\ensuremath{\mathcal{U}}}
\newcommand{\E}{\ensuremath{\mathcal{E}}}

\newcommand{\FK}{\ensuremath{\mathfrak{K}}}
\newcommand{\RgeO}{\ensuremath{\mathbb{R}_{ \geq 0}}}
\newcommand{\Rat}[1]{\ensuremath{\mathbb{R}^{#1}}}
\newcommand{\mbf}[1]{\ensuremath{\boldsymbol{#1}}} % % SOS NEW WAY FOR BOLD MATH SHOULD BE THE RIGHT WAY
\renewcommand{\bf}[1]{\textbf{#1}}

\begin{document}

\begin{frontmatter}

\title{Finite Horizon Discrete Models for Multi-Agent Control Systems with Coupled Dynamics} % Title, preferably not more

%\thanks[footnoteinfo]{This paper was not presented at any IFAC
%meeting. Corresponding author M.~T.~Cicero. Tel. +XXXIX-VI-mmmxxi.
%Fax +XXXIX-VI-mmmxxv.}

%\author[kth1]{Alexandros Nikou}\ead{anikou@kth.se},    % Add the
\author[ucsd1]{Dimitris Boskos}\ead{dboskos@ucsd.edu},               % e-mail 
%address
%\author[kth2]{Jana Tumova}\ead{tumova@kth.se},
\author[kth1]{Dimos V. Dimarogonas}\ead{dimos@kth.se}  % (ead) as shown

\address[ucsd1]{Department of Mechanical and Aerospace Engineering, 
University of California, San Diego.} 

\address[kth1]{School of Electrical Engineering 
and Computer Science, KTH Royal Institute of Technology, SE-100 44, 
Stockholm, Sweden.}  % Please supply
%\address[kth2]{School of Computer Science and Communication, KTH Royal Institute of Technology, SE-100 44, Stockholm, Sweden.}  % Please supply
% Five to ten keywords, chosen from the IFAC keyword list or with the help of the Automatica keyword wizard


% Abstract of not more than 200 words.

\begin{abstract}
%In this paper, we aim at the development of an online abstraction framework for multi-agent systems under coupled constraints. The motion capabilities of each agent are abstracted through a finite state transition system in order to capture reachability properties of the coupled multi-agent system over a finite time horizon in a decentralized manner. In the first part of this work, we define online abstractions by discretizing an overapproximation of the agents' reachable sets over the horizon. Then, sufficient conditions relating the discretization and the agent's dynamics properties are provided, in order to quantify the transition possibilities of each agent.  
The goal of this paper is to obtain online abstractions for coupled multi-agent 
systems in a decentralized manner. A discrete model which captures the motion 
capabilities of each agent is derived over a bounded time-horizon, by 
discretizing a corresponding overapproximation of the agent's reachable states. 
The individual abstractions' composition provides a correct representation of 
the coupled continuous system over the horizon and renders the approach 
appropriate for control synthesis under high-level specifications which are 
assigned to the agents over this time window. Sufficient conditions are also 
provided for the space and time discretization to guarantee the derivation of 
deterministic abstractions with tunable transition capabilities. 
\end{abstract}

\end{frontmatter}

\section{Introduction}

During the last decades there has been an emerging focus on the problem of 
high-level planning for  multi-agent systems by leveraging methods from formal 
verification, with applications in several domains such as 
robotics~\cite{LsKk04} and air traffic management~\cite{LsRdPa13}. In order to
exploit these tools for dynamic agents, it is required to build a discretized 
model of the continuous system, which allows for the algorithmic synthesis of 
high-level plans. Specifically, the use of an appropriate abstract 
representation enables the conversion of discrete paths into sequences of 
feedback controllers which enable the continuous-time model to implement the 
high-level specifications.
The aforementioned control synthesis problem has led to a significant  
research effort for the derivation of discrete state analogues of  continuous 
control systems, also called abstractions, which can capture reachability 
properties of the original model. Abstractions for piecewise affine 
systems on simplices and rectangles were introduced in  \cite{HlVj01} and have 
been further studied in \cite{BmGm14}. Closer related to the control framework 
that we adopt here for the derivation of 
the discrete models is the paper \cite{HmCp14b}, which builds on the notion of 
In-Block Controllability \cite{CpWy95}. According to this property, all 
pairs of states inside a block should be mutually reachable in finite time 
through a bounded control. Abstractions for nonlinear systems are derived in 
\cite{Rg11} based on convex overapproximations of reachable sets, and in 
\cite{AaTaSa09} for polynomial and other classes of systems. Other 
approaches build on approximate system relations~\cite{Tp09}, and have 
been exploited to derive abstractions for nonlinear stabilizable~\cite{Tp08}, 
incrementally forward complete~\cite{ZmPgMmTp12}, and networked 
control systems~\cite{ZmMmKmAa17}.

Reducing the complexity of the derived abstractions constitutes a 
significant research challenge. Towards this end, on the fly algorithms 
integrating the abstraction and control synthesis were considered in 
\cite{PgBaDm12}.
%rendering them appropriate for specifications confined on  
%parts of the whole workspace. 
In \cite{MsGaGg13}, multi-scale discrete models 
are introduced, whereas in
\cite{SeAa13},
%where finer discretization modes are only explored when 
%control  synthesis is no longer possible with coarser ones. 
local Lipschitz properties of probability densities  are considered to 
efficiently abstract stochastic systems into finite Markov 
Chains. 
%Using such local properties of the dynamics to obtain coarser models is 
%also a feature exploited in this paper.
%
Given the exponential complexity of the symbolic models  
with respect to the system's state 
dimension,  recent attention has been drawn on the 
consideration of abstractions for interconnected systems,  
based on discrete models for their building components.  
Towards this goal, compositional abstractions 
for stabilizable interconnected linear 
systems	were provided in \cite{TyIj08}. The work   
\cite{PgPpDm16} establishes approximate bisimulations for discrete time 
interconnected systems under small-gain assumptions.  
An algorithm suitable for system components with sparse interconnections is 
proposed in \cite{GfKeAm17}. Small-gain type assumptions are also leveraged in 
\cite{DeTp15}, whereas in \cite{MpGaWe15}, compositional 
abstractions for safety specifications are obtained for 
systems with monotone dynamics. In \cite{MkSzSaMr17l}, sufficient conditions 
are given to obtain bisimilar compositional abstractions for networks of 
incrementally input-to-state stable stochastic systems. Abstractions of 
feedback linearizable systems in cascade form were given in \cite{HoAaTp17}, 
and continuous abstractions in the form of lower dimensional control system 
models are derived in \cite{RmZm15}. Recent results focused more on 
verification consider also nonlinear coordinate transformations to provide 
decoupling abstractions for polynomial ODE's~\cite{SaGkJt16}.

Motivated by applications such as the coordination of robotic networks, 
we consider multi-agent systems evolving according to continuous-time 
dynamics. Assigning coupling feedback laws among appropriately nearby 
agents can provide an autonomous and scalable way to achieve collective 
objectives, including rendezvous to a common location and geometric 
formations while keeping the network connected and avoiding collisions 
\cite{MmEm10}. Additionally, it may be required to combine such 
objectives with high-level tasks, including safety, sequential 
reachability, and timed specifications \cite{NaBdTjDd18}. Such tasks can 
be assigned to the team or specific agents during runtime, and may need to be 
fulfilled within specific time bounds. Therefore, we consider agent dynamics
that apart from typical interconnection terms encountered in their popular 
coordination schemes include also bounded additive inputs, which provide some  
control capabilities to each agent. We exploit this control freedom to 
construct an individual transition system for each agent over a bounded time 
horizon that is used for correct-by-design  control synthesis. 
Despite of the available controllability, we only assume forward 
completeness for the dynamics and make no hypotheses about the network 
interconnection structure. This makes the compositional abstraction approaches 
discussed above in general not applicable to the systems we consider. 
Specifically, in~\cite{TyIj08}, \cite{MpGaWe15}, and~\cite{MkSzSaMr17l},  the 
dynamics need to be linear,  monotone, and incrementally input-to-state 
stable, respectively, which is not required here, whereas in~\cite{PgPpDm16}, 
the system components satisfy a small-gain condition that we do not impose. 
The works~\cite{DeTp15} and \cite{RmZm15} consider abstractions from 
discrete-state to discrete-state systems, and continuous-state to 
continuous-state systems, respectively, while in this paper, we focus on the 
continuous-state to discrete-state problem. Further, in~\cite{HoAaTp17}, the 
systems need to have a cascade interconnection, whereas here, the results are 
applicable for any interconnection structure.

The agents' abstractions are built over a fixed time horizon by 
discretizing an overapproximation of their corresponding reachable sets into 
cells and selecting a common time duration for their transitions. 
For appropriate space and time discretizations each agent can perform 
transitions from an initial cell to other ``nearby" ones by the selection of a 
feedback controller for each such transition. Selecting a common time step for 
the agents' abstractions enables the synchronization of their  
transitions and renders their product a correct representation 
of the coupled system. Thus, it is possible to synthesize correct-by-design 
control strategies for specifications that can be satisfied over the horizon by 
working exclusively at the discrete level. In addition, based on the agents' 
dynamics over the horizon, we specify acceptable discretizations which provide 
abstractions with quantifiable transition possibilities for each agent.
The approach builds in part on our recent works  
\cite{BdDd16} and \cite{BdDd19}, where the whole workspace is 
discretized, 
using global bounds for the agents' dynamics. The main contributions of 
the paper are summarized as follows. 
1) We derive deterministic agent models, which facilitates their algorithmic 
exploitation. This is accomplished by assigning decentralized feedback 
laws to the agents for the execution of their transitions, that attenuate an 
appropriate part of the coupling with their neighbors. Compared to our 
previous results \cite{BdDd16}, \cite{BdDd19}, which provide deterministic 
abstractions for analogous agent dynamics, we now only require that the system 
is forward complete. 2) We derive heterogeneous discretizations for the 
agents by using local bounds for their dynamics. This results in 
discretizations that are coarser, and hence, of reduced complexity for agents 
with higher actuation and weaker couplings over the horizon.
The abstractions are decentralized, in the sense that 
correctness of the individual discrete models' composition is not based on any 
global information on the interconnection structure. This further 
enables the identification of interconnection structures and specifications 
where control synthesis can be performed by avoiding the composition of all 
subsystems. Preliminary results from this paper appear in \cite{BdDd17a}. 
However, we include here all proofs which were omitted in \cite{BdDd17a}, an 
updated analysis, and a worked out example with simulations.

\section{Preliminaries and Notation}\label{sec:preliminaries}
We use the notation $|x|$ for the Euclidean norm of a vector $x\in\Rat{n}$. 
%       and denote the interior of a set $S\subset\Rat{n}$ by ${\rm int}(S)$. 
Given $R>0$ and $x\in\Rat{n}$, we define by $B(x;R):=\{y\in\Rat{n}:|x-y|\le R 
\}$ the closed ball with center $x\in\Rat{n}$ and radius $R$, and 
$B(R):=B(0;R)$. Given two sets $A,B\subset\Rat{n}$, their Minkowski sum is 
defined as $A+B:=\{x+y:x\in A, y\in B\}$. We say that $\beta:\RgeO\times\RgeO 
\to\RgeO$ is of class $\mathcal{K}_+\mathcal{K}_+$, if $\beta(t,\cdot)$ and 
$\beta(\cdot,s)$ are continuous and strictly increasing for all $t,s\ge 0$. 
Also, given $U\subset \Rat{n}$, we denote by $\M_U$ the set of measurable 
functions from $\RgeO$ to $U$ and define ${\rm sat}_r:\Rat{n}\to\Rat{n}$ as 
${\rm sat}_r(x)=x$ if $|x|\le r$ and ${\rm sat}_r(x)=r\frac{x}{|x|}$ if 
$|x|>r$, $r>0$. 

Given a multi-agent system with $N$ agents and an agent 
$i\in\N:=\{1,\ldots,N\}$ we use the notation 
$\mathcal{N}_i\subset\N\setminus\{i\}$ for the set of its neighbors, $N_i$ for 
its cardinality and define $\bar{\N}_i:=\N_i\cup\{i\}$. We consider an ordering 
of the agent's neighbors, denoted by $j_1,\ldots,j_{N_i}$ and define the 
$N_i$-tuple $j(i)=(j_1(i),\ldots,j_{N_i}(i))$. Whenever it is clear from the 
context, the argument $i$ will be omitted from the latter notation. The agents' 
network is represented by a directed graph $\G:=(\N,\E)$ with vertex set $\N$ 
the agents' index set and edge set $\E$ the ordered pairs $(\ell,i)$ with 
$i\in\N$ and $\ell\in\N_i$. A sequence $i_0i_1\cdots i_m$ with 
$(i_{\kappa-1},i_{\kappa})\in\E$, $\kappa=1,\ldots,m$, 
%namely, consisting of $m$ consecutive edges 
%in $\G$, forms \textit{a path} (of length $m$) in $\G$. A path $i_0i_1\cdots 
%i_m$ 
and $i_0=i_m$ is called a \textit{cycle}. Given the nonempty sets 
$\I_1,\ldots,\I_N$, their Cartesian product $\bfI:=\I_1\times\cdots\times\I_N$, 
and an agent $i\in\N$ with neighbors $j_1,\ldots,j_{N_i}\in\N$, the mapping 
${\rm pr}_i:\bfI\to\bfI_i:=\I_i\times\I_{j_1}\times\cdots\times\I_{j_{N_i}}$ 
assigns to each $N$-tuple $(l_1,\ldots,l_N)\in\bfI$ the $N_i+1$-tuple 
$(l_i,l_{j_1},\ldots,l_{j_{N_i}})\in\bfI_i$, i.e., the entries of agent $i$ and 
its neighbors. A transition system is a tuple 
$TS:=(Q,Q_0,Act,\longrightarrow)$, where: $Q$ is a set of states; 
$Q_0\subset Q$ is a set of initial states; $Act$ is a set of actions; 
$\longrightarrow$ is a transition relation with $\longrightarrow\subset Q\times 
Act\times Q$. The transition system is said to be finite, if $Q$ and $Act$ are 
finite sets. We also denote an element 
$(q,a,q')\in\longrightarrow$ as  $q\overset{a}{\longrightarrow} q'$ and define 
${\rm Post}(q;a):=\{q'\in Q:(q,a,q')\in\longrightarrow\}$, for every $q\in Q$ 
and $a\in Act$. $TS$ is called deterministic, if $|{\rm Post}(q;a)|\le 1$ 
for each $q\in Q$ and $a\in Act$, where $|\cdot|$ is used to denote the 
cardinality of a set. We say that $TS$ is nonblocking on $\bar Q\subset Q$, if 
for each $q\in\bar Q$, there exist $a\in Act$ with ${\rm 
Post}(q;a)\ne\emptyset$. Finally, given a domain $D\subset\Rat{n}$, a cell 
decomposition $\mathcal{S}=\{S_{l}\}_{l\in\mathcal{I}}$ of $D$ is 
a partition of $D$ into uniformly bounded sets with $S_l\cap 
S_{\hat l}=\emptyset$ for all $l\ne\hat l$, and $\cup_{l\in\mathcal{I}} 
S_{l}=D$. Given a set $A\subset D$, we say that $\S$ 
is compliant with $A$, if for any $S\in\S$ with $S\cap A\ne \emptyset$ it holds that $S\subset A$.

\section{Problem Formulation} \label{sec:pf}

In this section we provide the class of interconnected systems studied in 
the paper and formulate the decentralized abstraction problem that will be 
considered.   
We focus on multi-agent systems with dynamics
\begin{equation}\label{single:integrator}
\dot{x}_{i}(t)=f_{i}(x_{i}(t),\bf{x}_j(t))+v_{i}(t),x_{i}(t)\in \Rat{n},i\in\N,
\end{equation}
\noindent where 
$\bf{x}_j(t)(=\bf{x}_{j(i)}(t)):=(x_{j_{1}}(t),\ldots,x_{j_{N_i}}(t))\in\Rat{N_i
n}$. We consider decentralized control laws consisting of a feedback 
term $f_{i}(\cdot)$, which depends on the states of $i$ and its neighbors, and 
an additive input $v_{i}$ that provides certain control freedom to the agents 
and is leveraged for their abstractions. These dynamics can describe several 
multi-agent controllers, including consensus, connectivity maintenance, 
rendezvous, and formation~\cite{MmEm10}, which are encountered for instance in 
decentralized multi-robot coordination applications. Such interconnection 
protocols can also guarantee desired properties robustly with respect to the 
$v_i$'s, as for instance in \cite{BdDd17b}, where network 
connectivity is robustly preserved among neighboring agents for all additive  
inputs respecting a certain magnitude bound.
 Alternatively,~\eqref{single:integrator} can represent internal 
 dynamics of the system, as in temperature regulation of interconnected 
 rooms~\cite{AmDdShJk14}.  We will assume that each $f_{i}(\cdot)$ is 
 locally Lipschitz, and that 
 $v_{i}\in\mathcal{U}_{i}:=\M_{B(v_{\max}(i))}$, or equivalently, that
\begin{equation}\label{input:bound}
|v_{i}(t)|\le v_{\max}(i),\forall t\ge 0,
\end{equation}
for all $i\in\N$, and define 
$\mathcal{U}:=\M_{B(v_{\max}(1))\times\cdots\times B(v_{\max}(N))}$. In 
addition, we assume  that system \eqref{single:integrator} is \textit{forward 
complete}: 
\begin{assumption} \label{assumption:completeness}
\textbf{(Forward Completeness)} For every initial condition $\bf 
X_0\in\Rat{Nn}$ and input $\mbf v\in\mathcal{U}$ the solution $\bf 
x(t,\bf X_0;\mbf v)$ of~\eqref{single:integrator} is defined for all $t\ge 0$. 
Equivalently (see \cite[Lemma 2.3(ii)]{Ki05}), there exists a function 
$\beta\in\mathcal{K}_+\mathcal{K}_+$ 
such that 
for any  $\bf X_0\in\Rat{Nn}$ and $\mbf v\in\mathcal{U}$
\begin{equation} \label{beta:bound}
|\bf x(t,\bf X_0;\mbf v)|\le\beta (t,|\bf X_0|),\forall t\ge 0.
\end{equation}
\end{assumption}
%
%\subsection{Abstraction Requirements and Problem Statement}
%
Given any finite time horizon $[0,T]$ and the agents' positions 
at $t=0$, we will discretize an overapproximation $\R_i([0,T])$ of each 
agent's reachable set over $[0,T]$ and select a common transition time step 
$\delta t$ with $T=\ell\delta t$, $\ell\in\mathbb N$, to obtain its 
individual abstraction in a decentralized manner. We use the term 
\textit{decentralized} to indicate that the construction of each agent's 
symbolic model is based only on local network information. It is also required 
that the composition of the individual transition systems can capture correctly 
the behavior of the coupled system \eqref{single:integrator} over $[0,T]$.  
Assuming that a set of specifications which need to be fulfilled over a 
bounded time duration are assigned to the agents, we can select a horizon 
$[0,T]$ which is sufficient to infer about their satisfiability and leverage 
the abstractions to synthesize corresponding discrete plans. Such 
specifications can be timed reachability goals such as ``each agent $i$ should 
reach region $A(i)$ between $\tau_1(i)$ and $\tau_1'(i)$ units, and region 
$B(i)$ between $\tau_2(i)$ and $\tau_2'(i)$ units", with 
$0\le\tau_1(i)<\tau_1'(i)\le\tau_2(i)<\tau_2'(i)$, whose satisfiability 
can be checked over the finite time horizon $[0,T]$ with 
$T\equiv\max\{\tau_2'(i),i=1,\ldots,N\}$. Then, after selecting 
a satisfying plan and obtaining the agents' positions at the end of its 
execution, i.e., at $t=T$, the same procedure can be repeated for another task 
assignment over a new horizon $[T,T_1]$, by updating the abstractions, and 
analogously for subsequent horizons $[T_{\kappa},T_{\kappa+1}]$, $\kappa\ge 1$. 
This motivates also the term \textit{online} to characterize these abstractions.

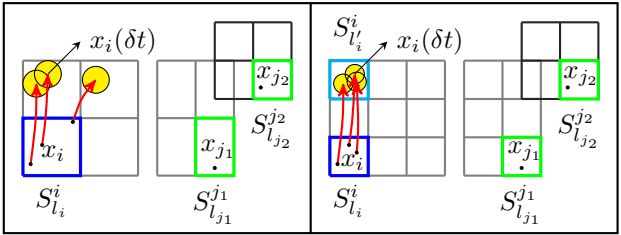
\begin{figure}[H]
\begin{center}
\begin{tikzpicture} [scale=.51]

% Agent i

\draw[color=gray,thick] (0,0) -- (3,0);
\draw[color=gray,thick] (0,1.5) -- (3,1.5);
\draw[color=gray,thick] (0,3) -- (3,3);

\draw[color=gray,thick] (0,0) -- (0,3);
\draw[color=gray,thick] (1.5,0) -- (1.5,3);
\draw[color=gray,thick] (3,0) -- (3,3);

\draw[color=blue,very thick] (0,0) -- (1.5,0) -- (1.5,1.5) -- (0,1.5) -- (0,0);

\fill[yellow] (.35,2.4) circle (0.35cm);
\fill[yellow] (.65,2.65) circle (0.35cm);
\fill[yellow] (1.9,2.5) circle (0.35cm);

\draw[black] (.35,2.4) circle (0.35cm);
\draw[black] (.65,2.65) circle (0.35cm);
\draw[black] (1.9,2.5) circle (0.35cm);

\draw [color=red,thick,->,>=stealth'](.2,0.3) .. controls (.35,1.5) .. (.35,2.4);
\fill[black] (.2,0.3) circle (1.5pt);

\draw [color=red,thick,->,>=stealth'](.5,0.8) .. controls (.6,1.5) .. (.65,2.65);
\fill[black] (.5,0.8) circle (1.5pt);

\draw [color=red,thick,->,>=stealth'](1.3,1.4) .. controls (1.5,2) .. (1.9,2.5);
\fill[black] (1.3,1.4) circle (1.5pt);

\draw[->,>=stealth] (.65,2.65) -- (1.5,3.5);
\coordinate [label=right:$x_{i}(\delta t)$] (A) at (1.5,3.5);

\coordinate [label=below:$S_{l_i}^i$] (A) at (0.75,0);
\coordinate [label=above left:$x_{i}$] (A) at (1.35,0.15);

% Agent j1

\draw[color=gray,thick] (3.5,0) -- (5.5,0);
\draw[color=gray,thick] (3.5,1.5) -- (5.5,1.5);
\draw[color=gray,thick] (3.5,3) -- (5.5,3);

\draw[color=gray,thick] (3.5,0) -- (3.5,3);
\draw[color=gray,thick] (4.5,0) -- (4.5,3);
\draw[color=gray,thick] (5.5,0) -- (5.5,3);

\draw[color=green,very thick] (4.5,0) -- (5.5,0) -- (5.5,1.5) -- (4.5,1.5) -- (4.5,0);

\coordinate [label=below:$S_{l_{j_1}}^{j_1}$] (A) at (5,0);
\fill[black] (5,.2) circle (1.5pt);
\coordinate [label=above:$x_{j_{1}}$] (A) at (5.1,.2);

% Agent j2

\draw[color=gray!30!black,thick] (5,2) -- (5,4);
\draw[color=gray!30!black,thick] (6,2) -- (6,4);
\draw[color=gray!30!black,thick] (7,2) -- (7,4);

\draw[color=gray!30!black,thick] (5,2) -- (7,2);
\draw[color=gray!30!black,thick] (5,3) -- (7,3);
\draw[color=gray!30!black,thick] (5,4) -- (7,4);

\draw[color=green,very thick] (6,2) -- (7,2) -- (7,3) -- (6,3) -- (6,2);

\coordinate [label=below:$S_{l_{j_2}}^{j_2}$] (A) at (6.5,2);
\fill[black] (6.2,2.3) circle (1.5pt);
\coordinate [label=above:$x_{j_{2}}$] (A) at (6.55,2.1);

%\coordinate [label=below:\red{Non well posed} discretization] (A) at (3,-2);

\draw[color=black,thick] (7.5,-1.5) -- (7.5,4.5) -- (-.5,4.5) -- (-.5,-1.5) -- (15.5,-1.5) -- (15.5,4.5) -- (7.5,4.5);
% % % % % % % % % % % %
% RIGTH DISCRETIZATION
% % % % % % % % % % % %

% Agent i

\draw[color=gray,thick] (8,0) -- (11,0);
\draw[color=gray,thick] (8,1) -- (11,1);
\draw[color=gray,thick] (8,2) -- (11,2);
\draw[color=gray,thick] (8,3) -- (11,3);

\draw[color=gray,thick] (8,0) -- (8,3);
\draw[color=gray,thick] (9,0) -- (9,3);
\draw[color=gray,thick] (10,0) -- (10,3);
\draw[color=gray,thick] (11,0) -- (11,3);

\draw[color=blue,very thick] (8,0) -- (9,0) -- (9,1) -- (8,1) -- (8,0);
\draw[color=cyan,very thick] (8,2) -- (9,2) -- (9,3) -- (8,3) -- (8,2);

\fill[yellow] (8.35,2.4) circle (0.25cm);
\fill[yellow] (8.65,2.65) circle (0.25cm);
\fill[yellow] (8.7,2.45) circle (0.25cm);

\draw[black] (8.35,2.4) circle (0.25cm);
\draw[black] (8.65,2.65) circle (0.25cm);
\draw[black] (8.7,2.45) circle (0.25cm);

\draw [color=red,thick,->,>=stealth'](8.2,0.3) .. controls (8.35,1.5) .. (8.35,2.4);
\fill[black] (8.2,0.3) circle (1.5pt);

\draw [color=red,thick,->,>=stealth'](8.5,0.8) .. controls (8.6,1.5) .. (8.65,2.65);
\fill[black] (8.5,0.8) circle (1.5pt);

\draw [color=red,thick,->,>=stealth'](8.7,0.6) .. controls (8.75,1.3) .. (8.7,2.45);
\fill[black] (8.7,0.6) circle (1.5pt);

\coordinate [label=below:$S_{l_i}^i$] (A) at (8.5,0);
\coordinate [label=above left:$x_{i}$] (A) at (9.15,-0.1);
\coordinate [label=above:$S_{l_i'}^i$] (A) at (8.5,3);

\draw[->,>=stealth] (8.65,2.65) -- (9.5,3.5);
\coordinate [label=right:$x_{i}(\delta t)$] (A) at (9.5,3.5);

% Agent j1

\draw[color=gray,thick] (11.5,0) -- (13.5,0);
\draw[color=gray,thick] (11.5,1) -- (13.5,1);
\draw[color=gray,thick] (11.5,2) -- (13.5,2);
\draw[color=gray,thick] (11.5,3) -- (13.5,3);

\draw[color=gray,thick] (11.5,0) -- (11.5,3);
\draw[color=gray,thick] (12.5,0) -- (12.5,3);
\draw[color=gray,thick] (13.5,0) -- (13.5,3);

\draw[color=green,very thick] (12.5,0) -- (13.5,0) -- (13.5,1) -- (12.5,1) -- (12.5,0);

\coordinate [label=below:$S_{l_{j_1}}^{j_1}$] (A) at (13,0);
\fill[black] (13,.2) circle (1.5pt);
\coordinate [label=above:$x_{j_{1}}$] (A) at (13.1,.15);

% Agent j2

\draw[color=gray!30!black,thick] (13,2) -- (13,4);
\draw[color=gray!30!black,thick] (14,2) -- (14,4);
\draw[color=gray!30!black,thick] (15,2) -- (15,4);

\draw[color=gray!30!black,thick] (13,2) -- (15,2);
\draw[color=gray!30!black,thick] (13,3) -- (15,3);
\draw[color=gray!30!black,thick] (13,4) -- (15,4);

\draw[color=green,very thick] (14,2) -- (15,2) -- (15,3) -- (14,3) -- (14,2);

\coordinate [label=below:$S_{l_{j_2}}^{j_2}$] (A) at (14.5,2);
\fill[black] (14.2,2.3) circle (1.5pt);
\coordinate [label=above right:$x_{j_{2}}$] (A) at (13.85,2.1);

%\coordinate [label=below:\blue{Well posed} discretization ] (A) at (11,-2);

\end{tikzpicture}
\caption{Illustration of space-time discretizations which are well posed (right) and non-well posed (left), respectively}. \label{figure:wellposed}
\end{center}
\end{figure}

For the space discretization we will consider cell decompositions 
$\{S_l^i\}_{l\in\I_i}$ of  the sets $\R_i([0,T])$, $i\in\N$. We denote by 
$\bf{l}_i=(l_i,l_{j_1},\ldots,l_{j_{N_i}})\in\bfI_i(=\I_i\times\I_{j_1}\times\cdots\times\I_{j_{N_i}})$
 the index tuple of the cells where agent $i$ and its neighbors belong and call 
it the \textit{cell configuration} of $i$. Given an analogous cell 
configuration $\bf{l}=(l_1,\ldots,l_N)\in\bfI(=\I_1\times\cdots\times\I_N)$ of 
all agents, we use the operator ${\rm pr}_i:\bfI\to\bfI_i$ from 
Section~\ref{sec:preliminaries} to obtain the cell configuration $\bf{l}_i={\rm 
pr}_i(\bf{l})$ of $i$. The state set of each agent's individual 
transition 
system is formed by the cells of its decomposition and its transition 
possibilities are affected only by the cells of its neighbors. In particular, 
given the cells of $i$ and its neighbors, a transition to a successor cell is 
enabled for $i$, if there exists a control law which can drive the agent to 
that cell at  $\delta t$, as in Fig.~\ref{figure:wellposed}, right. The control 
law needs to trigger this transition \textit{for all initial conditions of the 
agent and its neighbors in their cells, and irrespectively of the neighbors' 
further evolution during the transition interval}. For each agent, we consider 
also a set  $\R_i([0,T-\delta t])$ containing its reachable states over 
$[0,T-\delta t]$ and such that the trajectories of $i$ initiated from 
$\R_i([0,T-\delta t])$ remain in $\R_i([0,T])$ up to time $\delta t$. Thus, if 
a transition is initiated from a cell in $\R_i([0,T-\delta t])$, agent $i$ 
should reach some cell in $\R_i([0,T])$ at $\delta t$. We will require that 
each agent's transition system is \textit{nonblocking} and 
\textit{deterministic}, namely, that there exists an outgoing 
transition from each cell in $\R_i([0,T-\delta t])$, and each such transition 
is uniquely determined by the selected controller according to the transition 
requirement described above. A discretization which enables this property will 
be called \textit{well posed} (see Fig.~\ref{figure:wellposed}). The discussed  
abstraction requirements are summarized in 
the following problem formulation.

%This is illustrated in the right discretization of  Fig. 
%\ref{figure:wellposed}, where it is possible to drive  agent $i$ to cell 
%$S_{l_i'}^i$ at time $\delta t$ from any state in $S_{l_i}^i$, irrespectively 
%of its neighbors' initial positions in their cells and their corresponding 
%trajectories during the transition interval. Assuming that this holds for all 
%agents and their possible cell configurations, we obtain a well posed 
%discretization in this case. On the other hand, for the coarser discretization 
%on the left, the attainable sets of $i$ at $\delta t$ for the depicted initial 
%positions lie in different cells, implying that it is not well posed.
%
%
%\color{blue}

\noindent \textbf{Problem Formulation}
\textit{Given the time horizon $[0,T]$ and the agents' initial conditions 
$X_{10},\ldots,X_{N0}$, determine appropriate overapproximations 
$\R_i([0,t])$ of their reachable sets over $[0,t]$, $t\in[0,T-\tau]$, for 
some $\tau\in(0,T)$, select a cell decomposition $\{S_l^i\}_{l\in\I_i}$ of each 
$\R_i([0,T])$, and choose a common time step $\delta t<\tau$, in order to (i) 
derive a  deterministic transition system for each agent, which is nonblocking 
for all cells in $\R_i([0,T-\delta t])$; equivalently, establish that the 
discretization is well posed, and (ii) guarantee that any transition sequence 
in the composition of the agents' abstractions, which is initiated from cells 
containing their initial conditions $X_{10},\ldots,X_{N0}$, is matched by 
trajectories of system~\eqref{single:integrator} over the horizon that are 
generated by corresponding sequences of feedback controllers assigned to the 
agents.}

\section{Abstraction Domain and Transition Design} 
\label{sec:abstraction:basics}

In this section we start building towards the solution of the problem 
formulated above. In particular, we provide the overapproximations of the 
agent's reachable sets that will be discretized, and the control laws which 
are leveraged for their transitions. The purpose of each such controller is to 
attenuate part of the agent's coupling during a transition, so that the 
resulting trajectory is only dependent on the discrete location of its 
neighbors.
%	
%The agents' individual transition systems and appropriate discretizations for 
%the abstractions will be given in Section~\ref{sec:abstractions}, and their 
%correctness is shown in Section~\ref{sec:correctness:applicability}.
	
\subsection{Overapproximation of the Agents' Reachable Sets} \label{subsec:overapproximations}
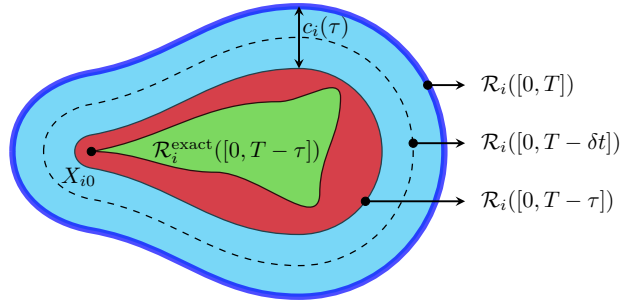
\begin{figure}[H]
\begin{center}

\resizebox{\columnwidth}{!}{
\begin{tikzpicture}[scale=.65]
\fill[fill=cyan!60!white, opacity=.7]  (-1.9,0) to[out=90,in=190] (0,1.9) to[out=10,in=180] (5,3.5) to[out=0,in=90] (8.5,0) to[out=-90,in=0] (5,-3.5) to[out=180,in=-10] (0,-1.9) to[out=170,in=-90] (-1.9,0);

\draw[line width=.02cm, dashed]   (-1.15,0) to[out=90,in=190] (0,1.14) to[out=10,in=180] (5,2.75) to[out=0,in=90] (7.75,0) to[out=-90,in=0] (5,-2.75) to[out=180,in=-10] (0,-1.15) to[out=170,in=-90] (-1.15,0);

\filldraw[line width=.02cm, fill=red, opacity=.7]  (-.4,0) to[out=90,in=190] (0,.4) to[out=10,in=180] (5,2) to[out=0,in=90] (7,0) to[out=-90,in=0] (5,-2) to[out=180,in=-10] (0,-.4) to[out=170,in=-90] (-.4,0);

%\filldraw[line width=.04cm, dashed, fill=red!40!white, opacity=.7]  (4,0) to[out=90,in=180] (5,2) to[out=0,in=90] (7,0) to[out=-90,in=0] (5,-2) to[out=180,in=-90] (4,0);

\filldraw[line width=.02cm, fill=green!60!white, opacity=.8]  (0,0) to[out=10,in=180] (4.5,1.2) to[out=0,in=90] (6,1.3) to[out=-90,in=90] (5.5,-1) to[out=-90,in=0] (3,-.5) to[out=180,in=0] (0,0);

\draw[line width=.1cm, color=blue, opacity=.7] (-1.9,0) to[out=90,in=190] (0,1.9) to[out=10,in=180] (5,3.5) to[out=0,in=90] (8.5,0) to[out=-90,in=0] (5,-3.5) to[out=180,in=-10] (0,-1.9) to[out=170,in=-90] (-1.9,0);

\filldraw[]  (0,0) circle (0.1cm);
\node (init) at (0.5,0) [label=below left:$X_{i0}$] {};

\node (init) at (3.5,0) [label=center:$\R_i^{\rm exact}({[0,T-\tau]})$] {};

\filldraw[]  (8.1,1.6) circle (0.1cm);
\draw[line width=.03cm,->,>=stealth] (8.1,1.6)  -- (9,1.6);

\node (init) at (9,1.6) [label=right:$\R_i({[0,T]})$] {};

\filldraw[]  (7.75,.2) circle (0.1cm);
\draw[line width=.03cm,->,>=stealth] (7.75,.2)  -- (9,.2);

\node (init) at (9,.2) [label=right:$\R_i({[0,T-\delta t]})$] {};

\filldraw[]  (6.6,-1.2) circle (0.1cm);
\draw[line width=.03cm,->,>=stealth] (6.6,-1.2)  -- (9,-1.2);

\node (init) at (9,-1.2) [label=right:$\R_i({[0,T-\tau]})$] {};

\draw[line width=.03cm,<->,>=stealth] (5,2)  -- (5,3.5);

\node (init) at (4.7,3) [label=right:$c_i(\tau)$] {};

\end{tikzpicture}
} %for resizebox
\end{center}
\caption{Illustration of the sets $\R_i([0,T-\tau])\subset\R_i([0,T-\delta 
t])$ $\subset\R_i([0,T])$, for $0<\delta t<\tau<T$. All contain the exact reachable set $\R_i^{\rm 
exact}([0,T-\tau])=\{x_i(t,\bf X_0;\mbf v):t\in[0,T-\tau]$, $\mbf v\in\U\}$ of 
$i$ over $[0,T-\tau]$ and its initial condition $X_{i0}$.}\label{fig:reach}
\end{figure}

For the subsequent analysis, we will assume fixed the initial states $X_{10},\ldots,X_{N0}$ of all agents at the beginning of the horizon $[0,T]$. 
%\color{blue}
Let 
\begin{equation} \label{iandneighbors:inclusion}
\R_{\bar{\N}_i}([0,T])\supset \{{\rm pr}_i(\bf x(t,\bf X_0;\mbf v)):t\in[0,T],\mbf v\in\U\}, 
\end{equation}
\noindent be an overapproximation of the states reachable by agent $i$ and its 
neighbors in $\Rat{(N_i+1)n}$ over the interval $[0,T]$ (recall that 
$\bar{\N}_i=\N_i\cup\{i\}$). Due to~\eqref{beta:bound}, each 
$\R_{\bar{\N}_i}([0,T])$ can be selected bounded, e.g., we can pick $\mathcal 
R_{\bar{\N}_i}([0,T])=\{(x_i,\bf x_j)\in\Rat{(N_i+1)n}:|(x_i,\bf 
x_j)|\le\beta(t,|\bf X_0|)\}$, with $\beta(\cdot)$ as given in (3). Hence, by 
continuity of the coupling terms there exist constants $M(i)>0$ such that
\begin{equation} \label{dynamics:bound}
|f_{i}(x_{i},\bf{x}_j)| \le M(i), \forall (x_{i},\bf{x}_j) \in \R_{\bar{\N}_i}([0,T]).
\end{equation}
Next, let $\tau\in(0,T)$ be an upper bound on the transition duration and 
consider bounded overapproximations $\R_i([0,T-\tau])\supset \{x_i(t,\bf 
X_0;\mbf v):t\in[0,T-\tau],\mbf v\in\U\}$ of each agent's reachable states in 
$\Rat{n}$ over $[0,T-\tau]$. Also, define 
\begin{align} 
c_i(\sigma):= & (M(i)+v_{\max}(i))\sigma, \label{cis:dfn} \\
\R_i([0,t]):= & \R_i([0,T-\tau])+B(c_i(t-T+\tau)).  \label{overapproximations:Ri} 
\end{align}
The sets $\R_i([0,t])$,  $t\in[T-\tau,T]$, depicted in 
Fig.~\ref{fig:reach}, 
are the overapproximations of the agents' reachable sets that will be 
used for 
the abstractions. Their key properties are provided in the following 
lemma, 
shown in the Appendix.
\begin{lemma} \label{lemma:overapproximations}
\textbf{(Overapproximation Properties)} For all $t\in[T-\tau,T]$, the sets 
$\R_i([0,t])$ in \eqref{overapproximations:Ri} satisfy 
\begin{align}
\R_i([0,t])+ & B(c_i(T-t))=\R_i([0,T]), \label{reach:set:equality} \\ 
\R_i([0,t])\supset & \{ x_i(s,\bf X_0;\mbf v):s\in[0,t],\mbf v\in\U\}. \label{reach:set:inclusion}
\end{align}
\end{lemma}
\color{black}
\subsection{Discrete Transition Control Design}

In order to derive the transitions of each agent $i\in\N$, we exploit the following auxiliary system 
\begin{equation} \label{system:disturbances}
\dot{z}_i(t)=g_i(z_i(t),\bf{d}_j(t))+v_i(t),
\end{equation}
\noindent with continuous disturbances 
$d_{j_1},\ldots,d_{j_{N_i}}:\RgeO\to\Rat{n}$. This approach is inspired by 
\cite{GaMs12}, where piecewise affine systems with disturbances are leveraged 
for the construction of hybridizations for nonlinear systems. These 
hybridizations are exploited to design robust controllers on simplices, whose 
objective, i.e., invariance control or control to a facet, is accomplished for 
all disturbances capturing the mismatch between the nonlinear model and the 
local affine approximations. Here the role of the disturbances is to capture 
the possible evolution of the agent's neighbors over the transition interval by 
restricting them over appropriate time-varying bounded sets. Each 
$g_i:\Rat{(N_i+1)n}\to\Rat{n}$ is given by
\begin{equation} \label{gi:dfn}
g_i(x_i,\bf{x}_j)={\rm sat}_{M(i)}(f_i(x_i,\bf{x}_j)),
\end{equation}
with $f_i(\cdot)$ and $M(i)$ as in \eqref{single:integrator} and \eqref{dynamics:bound}, respectively, and is locally Lipschitz. Thus, we get from \eqref{dynamics:bound} and \eqref{gi:dfn} that 
\begin{align} 
g_i(x_i,\bf{x}_j) & =f_i(x_i,\bf{x}_j),\forall (x_i,\bf{x}_j)\in \R_{\bar{\N}_i}([0,T]), \label{gi:extension} \\
|g_i(x_i,\bf{x}_j)| & \le M(i),\forall (x_i,\bf{x}_j)\in\Rat{(N_i+1)n}. \label{function:g} 
\end{align}
\noindent Recalling also that the sets $\R_i([0,T])$ have been selected 
bounded, there  exist constants  $L_1(i), L_2(i)>0$ with
\begin{align}
|g_i(x_i,\bf{x}_j) & -g_i(x_i,\bf{y}_j)|\le L_1(i)|\bf{x}_j-\bf{y}_j|, \label{dynamics:bound1} \\
|g_i(x_i,\bf{x}_j) & -g_i(y_i,\bf{x}_j)|\le L_2(i)|x_i-y_i|, \label{dynamics:bound2} \\
\forall x_{\kappa},y_{\kappa} & \in \R_{\kappa}([0,T]),\kappa\in\bar{\N}_i. \nonumber
\end{align}
The above bounds and Lipschitz constants of the vector fields $g_i(\cdot)$ will 
be used determine the size of acceptable space-time discretizations for the 
agents' abstractions over the horizon.

We next define the specific control laws that are exploited to derive the 
agents' transitions based on the auxiliary 
dynamics~\eqref{system:disturbances}. Consider for each agent $i$ a cell 
decomposition $\S_i=\{S_{l}^i\}_{l\in\I_i}$ of $\R_i([0,T])$ and a time step 
$\delta 
t$. We select the diameter $d_{\max}(i)$ of each $\S_i$ as the diameter of a 
ball whose translation can cover each cell in $\S_i$, and pick for every 
$S_{l_i}^i\in\S_i$ a reference point $x_{l_i,G}$ with 
\begin{equation} \label{reference:points}
|x_{l_i,G}-x|\le d_{\max}(i)/2,\forall x\in S_{l_i}^i.
\end{equation}

\begin{figure}[H]
\begin{center}

\resizebox{\columnwidth}{!}{
\begin{tikzpicture}[scale=1]

\filldraw[opacity=.2, color=magenta,thick] (3,0.5) circle (1cm);
\draw[color=magenta,thick] (3,0.5) circle (1cm);

\draw[color=gray] (0.1,0.8) -- (0.2,1.4) -- (1,1.5) -- (1.8,0.8);
\draw[color=gray] (1,1.5) -- (2.1,1.7);
\draw[color=gray] (0.3,-0.7) -- (0.8,0.1) -- (1.8,0.8)-- (1.6,-0.5) -- (0.3,-0.7);
\draw[color=gray] (0.1,0.8) -- (1.8,0.8);
\draw[color=green!50!black,thick] (-0.3,-0.5) -- (0.3,-0.7) -- (0.8,0.1) -- (0.1,0.8)-- (-0.3,-0.5);
\draw[color=cyan!70!black,thick] (1.8,0.8) -- (2.8,-0.7)-- (1.6,-0.5) -- (1.8,0.8);
\draw[color=cyan!70!black,thick] (1.8,0.8)-- (2.8,-0.7)-- (3.3,0.3) -- (3,1) -- (1.8,0.8);
\draw[color=cyan!70!black,thick] (2.8,-0.7)-- (4,-0.6) -- (4.1,0.4) -- (3.3,0.3) ;
\draw[color=cyan!70!black,thick] (1.8,0.8) -- (2.1,1.7) -- (3.2,1.7);
\draw[color=cyan!70!black,thick] (3,1) -- (3.2,1.7) -- (4.2,1.2) -- (4.1,0.4);

\fill[black] (3,0.5) circle (2pt);
\fill[black] (3.3,-0.2) circle (2pt);

\coordinate [label=left:$\textcolor{green!50!black}{S_{l_i}^i}$] (A) at (-0.3,-0.5);
\coordinate [label=right:$\textcolor{cyan!70!black}{S_{l_i'}^i}$] (A) at (3,-1);

\draw[->,>=stealth'] (3,0.5) -- (4.5,0.5) node[right] {$\chi_{i}(\delta t)$};
\draw[->,>=stealth'] (3.3,-0.2) -- (4.5,-0.2) node[right] {$z_{i}(\delta 
t)=\hat z_i(\delta t)=x$};

\filldraw[opacity=.2, color=magenta,thick] (6.1,.55) circle (.2cm);
\draw[color=magenta,thick] (6.1,.55) circle (.2cm);
\coordinate [label=right:$B(\chi_{i}(\delta t);r_i)$] (A) at (6.3,.5);

\draw[color=red,line width=.1cm] (7.2,1.2) -- (7.5,1.2);
\coordinate [label=right:$z_{i}(t)$] (A) at (7.5,1.2);

\draw[color=olive!50!black,line width=.1cm]  (5.9,1.2) -- (6.2,1.2);
\coordinate [label=right:$\hat{z}_{i}(t)$] (A) at (6.2,1.2);

\draw[color=blue,->,thick,>=stealth'] (0.2,0.05)  .. controls (1.4,0.7) .. (3,0.5);
\draw[color=olive!50!black,->,very thick,>=stealth'] (0.2,0.05)  .. controls (1.6,0.5) .. (3.3,-0.2);
\draw[color=red,->,thick,>=stealth'] (0.2,-0.6) .. controls (1.6,0.3) .. (3.3,-0.2);

\fill[black] (0.2,-0.6) circle (1.5pt) node[below] {$x_{i0}$};
\fill[black] (0.2,0.05) circle (2pt) node[below]{$x_{l_i,G}$};

\draw[color=black] (5.7,.2) -- (8.4,.2) -- (8.4,1.5) -- (5.7,1.5) -- (5.7,.2);

\end{tikzpicture}
}
\vspace{-0.4cm}

\end{center}
\caption{Illustration of the reference trajectory and reachability capabilities 
of the control laws \eqref{feedback:ki1}--\eqref{feedback:ki3}.} 
\label{fig:controllers}
\end{figure}
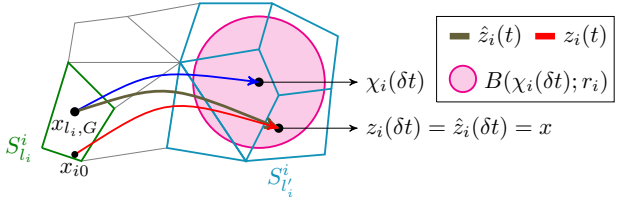

\noindent  For each agent $i$ and cell configuration $\bf{l}_i$ of $i$, 
define the feedback law  $k_{i,\bf{l}_i}:[0,\infty)\times 
\Rat{(N_i+1)n}\to\Rat{n}$, parameterized by $x_{i0}\in S_{l_i}^i$ and $w_i\in 
W_i$, as 
\begin{align}
k_{i,\bf{l}_i}(t,x_i,\bf{x}_j;x_{i0},w_i):= & {\rm 
sat}_{v_{\max}(i)}(\bar{k}_{i,\bf{l}_i}(t,x_i,\bf{x}_j;x_{i0},w_i)), 
\label{feedback:ki} \\
\bar{k}_{i,\bf{l}_i}(t,x_i,\bf{x}_j;x_{i0}, w_i):= & 
k_{i,\bf{l}_i,1}(t,x_i,\bf{x}_j) \nonumber \\
& + k_{i,\bf{l}_i,2}(w_i)+k_{i,\bf{l}_i,3}(x_{i0}) 
\label{feedback:ki:bar}
\end{align}
where
\begin{align}
W_i:= & B(v_{\max}(i))\subset\Rat{n}, \label{set:W} \\
k_{i,\bf{l}_i,1}(t,x_{i},\bf{x}_j):= & 
g_i(\chi_i(t),\bf{x}_{l_j,G})-g_i(x_i,\bf{x}_j), \label{feedback:ki1} \\
k_{i,\bf{l}_i,2}(w_i):= & \lambda(i)w_i, 
\lambda(i)\in[0,1),\label{feedback:ki2} \\
k_{i,\bf{l}_i,3}(x_{i0}):= & (x_{l_i,G}-x_{i0})/\delta t, 
\label{feedback:ki3} 
\end{align}
with $g_i(\cdot)$ as given in \eqref{system:disturbances}. The function  
$\chi_i(\cdot)$ in \eqref{feedback:ki1} is a reference trajectory defined 
through the solution of the initial value problem
\begin{equation} \label{reference:trajectory}
\dot{\chi}_i(t)=g_i(\chi_i(t),\bf{x}_{l_j,G}), \chi_i(0)=x_{l_i,G},
\end{equation}
where $x_{l_i,G}$, respectively, the $N_i$-tuple $\bf{x}_{l_j,G}$, represent 
the reference points of $i$ and its neighbors. The feedback laws 
$k_{i,\bf{l}_i}(\cdot)$ depend on the cell configuration $\bf{l}_i$ of agent 
$i$, through the reference points $\bf{x}_{l_j,G}$ and  $x_{l_i,G}$ in 
\eqref{feedback:ki1} and \eqref{feedback:ki3}, and the trajectory 
$\chi_i(\cdot)$ in \eqref{feedback:ki1} provided by 
\eqref{reference:trajectory}. Note that if the control law 
$v_i(t)=k_{i,\bf{l}_i,1}(t,z_{i}(t),\bf{d}_j(t))$ in~\eqref{feedback:ki1} is 
applied to agent $i$'s auxiliary dynamics~\eqref{system:disturbances} with 
$z_i(\cdot)$ initiated from $x_{l_i,G}$, then $z_i(\cdot)$ will coincide with 
the reference trajectory $\chi_i(\cdot)$. By adding also the component  
$k_{i,\bf{l}_i,2}(\cdot)$ in \eqref{feedback:ki2} to $v_i(t)$ 
in~\eqref{system:disturbances}, with $i$ initiated again from $x_{l_i,G}$, the 
agent will move according to the trajectory 
$t\mapsto\chi_i(t)+\lambda(i)w_it=:\hat z_i(t)$ and reach the point 
$x:=\chi_i(\delta t)+\lambda(i)w_i\delta t$ inside the ball in 
Fig.~\ref{fig:controllers} at $\delta t$. The point $x$ depends on the 
parameter $w_i$ from $W_i$ in \eqref{set:W}, whose selection enables the agent 
to reach any alternative point inside that ball. The ball's radius is
\begin{equation} \label{distance:r}
r_i:=\lambda(i)\delta tv_{\max}(i),
\end{equation}
and is tuned through the parameter  $\lambda(i)$ in~\eqref{feedback:ki2}, 
namely, the input part exploited to enable multiple transitions 
from each cell.  By leveraging also the component 
$k_{i,\bf{l}_i,3}(\cdot)$, i.e., by applying 
the unsaturated version $\bar k_{i,\bf{l}_i}(\cdot)$ of the feedback law, the 
agent can reach any point inside this ball from every initial condition in its 
cell, and perform a transition to any cell intersecting 
$B(\chi_i(\delta t);r_i)$. This is shown in the following lemma.
\begin{lemma} \label{lemma:unsaturated:control:trajectory}
\textbf{(Trajectories of \eqref{system:disturbances} Under Unsaturated 
Feedback)} Consider a cell 
configuration 
$\bf l_i$, the reference trajectory $\chi_i(\cdot)$ in 
\eqref{reference:trajectory}, and $r_i$ as given by \eqref{distance:r}. Then, 
(i) for any $x_{i0}\in S_{l_i}^i$, $w_i\in W_i$, and continuous disturbances 
$\bf d_j(\cdot)$, the solution $z_i(\cdot)\equiv z_i(\cdot;x_{i0},w_i)$ 
of~\eqref{system:disturbances} with 
initial condition $z_i(0)=x_{i0}$ and $v_i(t)=\bar k_{i,\bf 
l_i}(t,z_i(t),\bf d_j(t);x_{i0},w_i)$ with $\bar k_{i,\bf 
l_i}(\cdot)$ as in \eqref{feedback:ki:bar}, satisfies  
\begin{equation}  \label{xi:trajectory}
z_i(t) = \frac{\delta t-t}{\delta 	
t}(x_{i0}-x_{l_i,G})+\lambda(i)w_it+\chi_i(t), t\ge 0;
\end{equation}
(ii) for any $x\in B(\chi_i(\delta t);r_i)$, if we set 
\begin{equation}  \label{vector:wi}
w_i:= (x-\chi_i(\delta t))/(\lambda(i)\delta t),
\end{equation}
then $w_i\in W_i$ and $z_i(\delta t;x_{i0},w_i)=x$ for all $x_{i0}\in 
S_{l_i}^i$.
\end{lemma}
\begin{pf}
To prove (i), we get from \eqref{feedback:ki:bar} and 
\eqref{feedback:ki1}--\eqref{reference:trajectory}, that the 
solution of \eqref{system:disturbances} with $z_i(0)=x_{i0}$ and 
$v_i(t)=\bar k_{i,\bf{l}_i}(t,z_i(t),\bf{d}_j(t);x_{i0},w_i)$ is given as 
$z_i(t)=x_{i0}+\int_0^t(g_i(\chi_i(s),\bf{x}_{l_j,G})+\frac{1}{\delta 
t}(x_{l_i,G}-x_{i0})+\lambda(i) w_i)ds = 
x_{i0}+\frac{t}{\delta 	t}(x_{l_i,G}-x_{i0})+\lambda(i) 
w_it +\int_0^tg_i(\chi_i(s),\bf{x}_{l_j,G})ds = \frac{\delta 
t-t}{\delta t}(x_{i0}-x_{l_i,G}) 
+\lambda(i)w_it+x_{l_i,G}+\int_0^tg_i(\chi_i(s),\bf{x}_{l_j,G})ds=
 \frac{\delta t-t}{\delta 
t}(x_{i0}-x_{l_i,G})+\lambda(i)w_it+\chi_i(t)$. Hence, 
\eqref{xi:trajectory} holds. To show (ii), note that due to \eqref{distance:r} 
and \eqref{vector:wi}, $|w_i|\le \frac{r_i}{\lambda(i)\delta t}\le 
v_{\max}(i)$, and hence, by virtue of \eqref{set:W}, $w_i\in W_i$.
In addition, we get from \eqref{xi:trajectory} 
that $z_i(\delta t)=\chi_i(\delta t)+\lambda(i)w_i\delta t$, and thus, by 
\eqref{vector:wi}, that $z_i(\delta t)=x$. 
\end{pf}
\color{black}
The above result remains is fact also valid when the control law $k_{i,\bf 
l_i}$ in \eqref{feedback:ki} is applied, as long as the discretization 
guarantees that the saturation level of the controller is not exceeded. Such 
discretizations will be provided in Section~\ref{sec:abstractions}. It will 
also be shown in the correctness results of 
Section~\ref{sec:correctness:applicability} that such 
transitions can be performed from appropriate cell configurations by assigning 
these control laws to the coupled agents in \eqref{single:integrator}.

\section{Well-Posed Decentralized Abstractions} \label{sec:abstractions}

Here, we build on the previous section to define the agents' 
individual discrete models and provide space and time discretizations 
which lead to a deterministic and nonblocking transition system for each agent.

\subsection{Individual Discrete Agent Models}

To derive the transitions of each agent's discrete model we leverage the 
system with disturbances \eqref{system:disturbances} and control laws as 
the ones introduced in~\eqref{feedback:ki}. The results are presented in 
a more general context, in order to focus on the desirable properties of the 
control laws irrespectively of their precise formulas and keep the flexibility 
of alternative control designs to the one in \eqref{feedback:ki}. We consider 
hybrid feedback laws  parameterized by the agents' initial conditions and a set 
of auxiliary parameters associated to the agent's reachability 
capabilities. In particular, consider an agent $i\in\N$, cell decompositions  
$\S_{\kappa}=\{S_l^{\kappa}\}_{l\in\I_{\kappa}}$ of $\R_{\kappa}([0,T])$, 
$\kappa\in\bar{\N}_i(=\N_i\cup\{i\})$, a nonempty subset $W_i$ of $\Rat{n}$, 
and a cell configuration $\bf{l}_i$ of $i$. A control law  
$k_{i,\bf{l}_i}(t,x_i,\bf{x}_j;x_{i0},w_i):[0,\infty)\times \Rat{(N_i+1)n}\to 
S\subset\Rat{n}$, parameterized by $x_{i0}\in S_{l_i}^i$ and $w_i\in 
W_i$, which is locally Lipschitz continuous on $(x_i,\bf{x}_j)$, will be 
called a $W_i$ feedback law, and a $v_{\max}(i)-W_i$ feedback law if its 
codomain satisfies $S=B(v_{\max}(i))$. In the sequel, whenever we 
refer to a space-time discretization $\{\S_i\}_{i\in\N}-\delta t$ (or 
$\{\S_{\kappa}\}_{\kappa\in\bar{\N}_i}-\delta t$ for some $i$), it will 
be implicitly assumed that each $\mathcal{S}_i=\{S_l^i\}_{l\in\I_i}$ is a 
cell decomposition of $\R_i([0,T])$, and, that $\delta t<\tau$ and 
$T=\ell\delta t$ for some $\ell\in\mathbb N$. The following definition 
formalizes each agent's transition requirement, based on the auxiliary 
system~\eqref{system:disturbances} and knowledge of its neighbors' discrete 
positions.

\begin{dfn} \label{conditionCC}
\textbf{(Transition Requirement)} Consider an agent $i\in\N$, a space-time 
discretization $\{\S_{\kappa}\}_{\kappa\in\bar{\N}_i}-\delta t$, a cell 
configuration $\bf{l}_i$ of $i$ with $S_{l_{\kappa}}^{\kappa}\subset 
\R_{\kappa}([0,T-\delta t])$ for all $\kappa\in\bar{\N}_i$, and a $W_i$ 
feedback law $v_i=k_{i,\bf{l}_i}(t,x_{i},\bf{x}_j;x_{i0},w_i)$. Given $w_i\in 
W_i$ and a cell index $l_i'\in\I_i$, we say that $k_{i,\bf{l}_i}$, $w_i$, 
$l_i'$ satisfy the Transition Requirement (TR) if for each $x_{i0}\in 
S_{l_i}^i$ and disturbances $d_{\kappa}\in C(\RgeO;\Rat{n})$, $\kappa\in\N_i$, 
with 
\begin{equation}
d_{\kappa}(t) \in S_{l_{\kappa}}^{\kappa}+B(c_{\kappa}(t))  \label{disturbance:bounds}
\end{equation}
\noindent for all $t\in [0,\delta t]$, the solution $z_i(\cdot)$ of 
\eqref{system:disturbances} with 
$v_i(t)=k_{i,\bf{l}_i}(t,z_i(t),\bf{d}_j(t);x_{i0},w_i)$ and  
$z_i(0)=x_{i0}$, satisfies $z_i(\delta t) \in S_{l_i'}^i$, where  
$c_{\kappa}(\cdot)$ is given in \eqref{cis:dfn}.
\end{dfn}
Note that according to the Transition Requirement, agent $i$ can 
perform a deterministic transition  
to cell $S_{l_i'}^i$ precisely at $\delta t$ under the auxiliary 
dynamics \eqref{system:disturbances} using the feedback law 
$k_{i,\bf{l}_i}(\cdot)$ tuned by the corresponding parameter $w_i$. This 
is possible for all disturbances satisfying \eqref{disturbance:bounds}, which 
capture the evolution of $i$'s neighbors on $[0,\delta t]$. We next define 
well-posed discretizations for the agents, which guarantee that their 
abstractions are nonblocking and the Transition Requirement is satisfied by 
control laws respecting the bounds $v_{\max}(i)$.
\begin{dfn}\label{dfn:well:posed:discretization}
\textbf{(Well-Posed Discretizations)} Consider the space-time discretization 
$\{\S_i\}_{i\in\N}-\delta t$
\textit{(i)} Given an agent $i\in\N$, we say that 
$\{\S_{\kappa}\}_{\kappa\in\bar{\N}_i}-\delta t$ is well posed for~$i$, if for 
any initial cell configuration $\bf{l}_i\in\bfI_i$ with 
$S_{l_{\kappa}}^{\kappa}\subset\R_{\kappa}([0,T-\delta t])$ for all 
$\kappa\in\bar{\N}_i$, there exist a $v_{\max}(i)-W_i$ feedback law 
$k_{i,\bf{l}_i}$, a vector $w_i\in W_i$, and a cell index $l_i'\in\I_i$, which 
satisfy the Transition Requirement of Definition~\ref{conditionCC}.
\textit{(ii)} We say that the space-time discretization 
$\{\S_i\}_{i\in\N}-\delta t$ is well posed, if 
$\{\S_{\kappa}\}_{\kappa\in\bar{\N}_i}-\delta t$ is well 
posed for each $i\in\N$.
\end{dfn}
Following the formalism in~\cite{BdDd19} to determine the actions of 
the 
transitions, consider a well-posed discretization $\{\S_i\}_{i\in\N}-\delta t$. 
Also, select for each $i$ and  cell configuration $\bf{l}_i\in\bfI_i$ with 
$S_{l_{\kappa}}^{\kappa}\subset\R_{\kappa}([0,T-\delta t])$,  
$\kappa\in\bar{\N}_i$, a $v_{\max}(i)-W_i$ feedback law 
$k_{i,\bf l_i}$, generating at least one transition, namely, so that $k_{i,\bf 
l_i}$, $w_i$, $l_i'$ satisfy the Transition 
Requirement for some $w_i\in W_i$, $l_i'\in\I_i$, and define for all $l\in\I_i$
\begin{align} \label{wi:equiv:class}
[w_i]_{(\bf l_i,l)}:=\{w_i\in W_i:k_{i,\bf{l}_i},w_i,l\;\textup{satisfy (TR)}\}.
\end{align}
The sets $[w_i]_{(\bf l_i,l)}$ represent the parameters in $W_i$ which enable 
a transition to cell $S_l^i$ under the control $k_{i,\bf{l}_i}$ and are 
used to define the agents' individual abstractions. 
 
\begin{dfn} \label{individual:ts}
\textbf{(Individual Transition Systems)}  Consider a well-posed discretization 
$\{\S_i\}_{i\in\N}-\delta t$, an agent $i\in\N$, and for each cell 
configuration $\bf{l}_i\in\bfI_i$ with 
$S_{l_{\kappa}}^{\kappa}\subset\R_{\kappa}([0,T-\delta t])$,  	
$\kappa\in\bar{\N}_i$, a $v_{\max}(i)-W_i$ feedback law $k_{i,\bf l_i}$ which 
generates at least one transition. The individual transition system 
$TS_i:=(Q_i,Q_{0i},Act_i,$ $\longrightarrow_i)$ of agent $i$ consists of: State 
set $Q_i:=\I_i$ (the cell indices of $\S_i$); Initial state set 
$Q_{0i}:=\{l_i\in\I_i:X_{i0}\in S_{l_i}^i\}$; Actions $Act_i:=\bfI_i\times 
2^{W_i}$; Transition relation $\longrightarrow_i\subset Q_i\times 
Act_i\times Q_i$ defined as follows. For any $l_i,l_i'\in Q_i$ and 
$\bf{l}_i=(l_i,l_{j_1},\ldots,l_{j_{N_i}})\in\bfI_i$, 
$l_i\overset{(\bf{l}_i,[w_i])}{\longrightarrow_i}l_i'$ iff 
$S_{l_{\kappa}}^{\kappa}\subset\R_{\kappa}([0,T-\delta t])$,  
$\kappa\in\bar{\N}_i$, $[w_i]=[w_i]_{(\bf l_i,l_i')}$, and $[w_i]_{(\bf 
l_i,l_i')}\ne\emptyset$. Also, define
\begin{align}
\overline{{\rm Post}}_i(\bf l_i) & := \cup_{[w_i]\in 2^{W_i}}{\rm 
Post}_i(l_i;(\bf{l}_i,[w_i])) \nonumber  \\
= & \{l_i'\in\I_i:\,\exists\,[w_i]\in 2^{W_i}{\rm 
s.t.}\;l_i\overset{(\bf{l}_i,[w_i])}{\longrightarrow_i}l_i'\}, \label{Post:bar}
\end{align} 
providing all successor cells from configuration  $\bf l_i$. 
\end{dfn}

\begin{remark}
(i) Given a well-posed discretization, it follows from 
Definition~\ref{individual:ts} that the associated transition system $TS_i$ of 
each agent is nonblocking on $\bar 	
Q_i=\{l\in\I_i:S_l^i\subset\R_i([0,T-\delta t])\}$. In particular,  
transitions from cell configurations with 
$S_{l_{\kappa}}^{\kappa}\subset\R_{\kappa}([0,T-\delta t])$,  
$\kappa\in\bar{\N}_i$ are considered, which suffices in order to capture the 
agents' behavior up to the horizon end.
%In fact the transitions of 
%possible interest are the ones where these cells intersect the agents' exact 
%reachable sets over $[0,T-\delta t]$. Since the latter cannot be computed 
%explicitly in principle, we require that these cells are contained in the 
%reachable sets' overapproximations over $[0,T-\delta t]$. 
(ii) From \eqref{wi:equiv:class} and Definition~\ref{individual:ts}, it follows 
that each $[w_i]$ in the action of a transition is uniquely determined by 
the successor cell. Thus, by recalling that a cell 
decomposition is also a partition, each $TS_i$ is deterministic.
\end{remark}

\subsection{Space and Time Discretizations}

We next present sufficient conditions to obtain well-posed space-time 
discretizations for system \eqref{single:integrator}. Since the abstraction is 
updated at the end of the horizon, it is convenient to consider different 
decomposition diameters for the agents. 
This enables us to derive coarser discretizations for agents with weaker 
couplings and less conservative control constraints, i.e., larger 
$v_{\max}(i)$, and hence, reduce the abstraction complexity.   
We therefore introduce for each agent the design constraint that the 
diameters of its neighbors' decompositions satisfy
\begin{equation} \label{diameters:restrictions}
d_{\max}(j)\le \mu(j,i) d_{\max}(i).
\end{equation}
\noindent For these restrictions to be feasible, we require that 
\begin{align}
& \mu(i_0,i_1)\mu(i_1,i_2)\cdots\mu(i_{m-1},i_m)\ge 1,\nonumber \\
& \textup{for all cycles} \;i_0i_1\cdots i_m\;{\rm with}\;i_0=i_m \;{\rm in}\; 
\mathcal{G}, \label{diameters:restrictions:consistency}
\end{align}
\noindent which is always satisfied if we select $\mu(j,i)=1$ for all $i\in\N$ 
and $j\in\N_i$. For the acceptable values of the discretizations, it is also 
convenient to define for each agent the following local network parameters 
\begin{align}
\mbf{\mu}(i):= & (\sum_{j\in\N_i}\mu(j,i)^2)^{\frac{1}{2}}, 
\label{constant:mui}\\
\bf{M}(i):= &  (\sum_{j\in\N_i}(M(j)+v_{\max}(j))^2)^{\frac{1}{2}}. 
\label{constant:bfM(i)}
\end{align}
To obtain the agents' transitions according to Definition~\ref{individual:ts} 
we will use the control laws introduced in~\eqref{feedback:ki} and show that 
they do not exceed their saturation level during the transitions. We therefore 
derive bounds on the unsaturated components of~\eqref{feedback:ki} along the 
solutions of~\eqref{system:disturbances} in the following lemma, shown in the 
Appendix. 
%
%\color{blue}
%
\begin{lemma} \label{lemma:bounds}
\textbf{(Bounds on Control Components)} Consider a discretization  
$\{\S_i\}_{i\in\N}-\delta t$ with decomposition diameters $d_{\max}(i)$,  
parameters $\lambda(i)\in[0,1)$, $\mu(j,i)\ge 0$, $j\in\N_i$, and assume 
that \eqref{diameters:restrictions} and 
\eqref{diameters:restrictions:consistency} hold. Given $i\in\N$, a cell 
configuration $\bf{l}_i$ of $i$ with 
$S_{l_{\kappa}}^{\kappa}\subset\R_{\kappa}([0,T-\delta t])$ for all 
$\kappa\in\bar{\N}_i$, and continuous $d_{j_1},\ldots,d_{j_{N_i}}$ 
satisfying  \eqref{disturbance:bounds}, denote by $z_i(\cdot)$ the 
solution of~\eqref{system:disturbances} with $v_i=k_{i,\bf l_i}$ as 
in~\eqref{feedback:ki}. Then, for each $x_{i0}\in S_{l_i}^i$ and $w_i\in 
W_i$, the components $k_{i,\bf{l}_i,1}(\cdot)$, $k_{i,\bf{l}_i,2}(\cdot)$, 
and $k_{i,\bf{l}_i,3}(\cdot)$ in~\eqref{feedback:ki:bar} satisfy 
\begin{align}
|k_{i,\bf{l}_i,1}(t,x_{i}(t),\bf{d}_j(t))| & \le 
L_{1}(i)\left(\mbf{\mu}(i)\frac{d_{\max}(i)}{2}+\bf{M}(i)t\right) \nonumber \\
+L_{2}(i)|z_i(t) & -\chi_i(t)|, \forall t\in[0,\delta t], 
\label{ki1:bound} \\
|k_{i,\bf{l}_i,2}(w_i)| & \le \lambda(i) v_{\max}(i), 
\label{ki2:bound} \\
|k_{i,\bf{l}_i,3}(x_{i0})| & \le 
d_{\max}(i)/(2\delta t). \label{ki3:bound} 
\end{align}
with $L_1(i)$, $L_2(i)$, and $v_{\max}(i)$ as given in 
\eqref{dynamics:bound1}, \eqref{dynamics:bound2}, and \eqref{input:bound}, 
respectively,  $\mbf{\mu}(i)$, $\bf{M}(i)$ as defined in 
\eqref{constant:mui}, \eqref{constant:bfM(i)}, and the reference trajectory 
$\chi_i(\cdot)$ provided by \eqref{reference:trajectory}. 
\end{lemma}
Based on this lemma, we characterize discretizations where the controllers 
designed for the transitions do not exceed their saturation level over the 
transition interval.   
\begin{prop}\label{prop:discretizations:keep:k:unsaturated}
\textbf{(Discretizations Keeping Controller Below Saturation)} 
Consider a discretization $\{\S_i\}_{i\in\N}-\delta t$ with decomposition 
diameters $d_{\max}(i)$, and parameters $\lambda(i)\in[0,1)$, $\mu(j,i)\ge 
0$, $j\in\N_i$. Assume that \eqref{diameters:restrictions}, 
\eqref{diameters:restrictions:consistency} hold, and that $d_{\max}(i)$, 
$\delta t$ satisfy:
\begin{align}
& \delta t \in 
\left(0,\frac{(1-\lambda(i))v_{\max}(i)}{L_1(i)\bf{M}(i)+L_2(i)\lambda(i)v_{\max}(i)}\right),
\label{deltat:interval} \\
& d_{\max}(i) \in \left(0,\min\left\lbrace 
\tfrac{2(1-\lambda(i))v_{\max}(i)\delta 
t}{1+(L_1(i)\mbf{\mu}(i)+L_2(i))\delta t},\right.\right. \nonumber \\
& \hspace{3em} 
\left.\left.\tfrac{2(1-\lambda(i))v_{\max}(i)\delta 
t-2(L_1(i)\bf{M}(i)+L_2(i)\lambda(i)v_{\max}(i)) \delta 
t^2}{1+L_1(i)\mbf{\mu}(i)\delta t} \right\rbrace\right), 
\label{dmax:interval}
\end{align}
\noindent with $L_1(i)$, $L_2(i)$, and $v_{\max}(i)$ as given in 
\eqref{dynamics:bound1}, \eqref{dynamics:bound2}, and \eqref{input:bound}, 
respectively,  and $\mbf{\mu}(i)$, $\bf{M}(i)$ as defined in 
\eqref{constant:mui}, \eqref{constant:bfM(i)}. Also, let any $i\in\N$ and a 
cell configuration $\bf{l}_i$ of $i$ with 
$S_{l_{\kappa}}^{\kappa}\subset\R_{\kappa}([0,T-\delta t])$ for all 
$\kappa\in\bar{\N}_i$. Then, for any $x_{i0}\in S_{l_i}^i$, $w_i\in W$ and 
continuous $d_{j_1},\ldots,d_{j_{N_i}}$ satisfying  \eqref{disturbance:bounds}, 
the solution $z_i(\cdot)$ of \eqref{system:disturbances} with 
$v_i(t)=k_{i,\bf{l}_i}(t,z_i(t),\bf{d}_j(t);x_{i0},w_i)$ and		
$z_i(0)=x_{i0}$ satisfies 
$k_{i,\bf{l}_i}(t,z_i(t),\bf{d}_j(t);x_{i0},w_i)=\bar 
k_{i,\bf{l}_i}(t,z_i(t),\bf{d}_j(t);x_{i0},w_i)$ for all $t\in[0,\delta 
t]$, with $k_{i,\bf l_i}$  and $\bar k_{i,\bf l_i}$ as given by 
\eqref{feedback:ki} and \eqref{feedback:ki:bar}, respectively. 
\end{prop}
The proof of Proposition~\ref{prop:discretizations:keep:k:unsaturated} is given 
in the Appendix. We next provide the desired sufficient conditions for well 
posed space-time discretizations together with the agents' corresponding 
transition capabilities.
\begin{thm}\label{discretizations:for:planning}
\textbf{(Well-Posed Discretizations: Sufficient Conditions)} Consider a 
discretization $\{\S_i\}_{i\in\N}-\delta t$ with decomposition diameters 
$d_{\max}(i)$, and assume that the hypotheses of 
Proposition~\ref{prop:discretizations:keep:k:unsaturated} are fulfilled. Then, 
the space-time discretization is well posed for system 
\eqref{single:integrator}. In particular, by selecting for each $i\in\N$ and 
cell configuration $\bf{l}_i$ of $i$ with 
$S_{l_{\kappa}}^{\kappa}\subset\R_{\kappa}([0,T-\delta t])$,  
$\kappa\in\bar{\N}_i$, the control law $k_{i,\bf l_i}$ in \eqref{feedback:ki}, 
it holds that $\overline{{\rm Post}}_i(\bf l_i)\ne\emptyset$, with 
$\overline{{\rm Post}}_i(\cdot)$ as defined in \eqref{Post:bar}. 
Furthermore,     
\begin{align}
& B(\chi_i(\delta t);r_i) \subset \R_i([0,T]), \label{reachball:inclusion}\\
& \overline{{\rm Post}}_i(\bf l_i)=\{l\in\I_i:S_l^i\cap B(\chi_i(\delta 
t);r_i)\ne\emptyset\}, \label{planning:condition}
\end{align}
\noindent with $\chi_i(\cdot)$ and $r_i$ as given by 
\eqref{reference:trajectory} and \eqref{distance:r}, respectively.
\end{thm}

\begin{pf}
For the proof, pick $i\in\N$, and a cell configuration $\bf{l}_i$ of $i$ 
with $S_{l_{\kappa}}^{\kappa}\subset\R_{\kappa}([0,T-\delta t])$, 
$\kappa\in\bar{\N}_i$. Also, consider the reference trajectory 
$\chi_i(\cdot)$ in \eqref{reference:trajectory} which is defined for all 
$t\ge 0$.
We first show that \eqref{reachball:inclusion} is satisfied. Indeed, let 
any $x\in B(\chi_i(\delta t);r_i)$. Then, we obtain from \eqref{distance:r} 
that $|\chi_i(\delta t)-x|\le\lambda(i)v_{\max}(i)\delta t$. Furthermore, we 
get from \eqref{function:g} and \eqref{reference:trajectory} that 
$|\chi_i(\delta t)-x_{l_i,G}|\le M(i)\delta t$. Thus, it follows from 
\eqref{cis:dfn} that $x-x_{l_i,G}\in B(c_i(\delta t))$. From the latter, 
\eqref{reach:set:equality}, which implies that $\R_i([0,T-\delta 
t])+B(c_i(\delta t))=\R_i([0,T])$, and the fact that 
$x_{l_i,G}\in\R_i([0,T-\delta t])$, we obtain that $x\in\R_i([0,T])$, which 
establishes \eqref{reachball:inclusion}. 

Next, from \eqref{reachball:inclusion} and the fact that $\cup_{l_i\in\I_i} 
S_{l_i}^i=\R_i([0,T])$, we get that $\{l\in\I_i:S_l^i\cap 
B(\chi_i(\delta t);r_i)\ne\emptyset\}\ne\emptyset$. Thus, validity of 
\eqref{planning:condition} implies directly that $\overline{{\rm Post}}_i(\bf 
l_i)\ne\emptyset$. To show \eqref{planning:condition}, note first that due to 
\eqref{wi:equiv:class} and Definition~\ref{individual:ts}, $\overline{{\rm 
Post}}_i(\bf l_i)=\{l\in\I_i:\exists\, w_i\in W_i\;\textup{s.t.}\;k_{i,\bf 
l_i},w_i,l\;{\rm satisfy}\;{\rm (TR)}\}$. Thus, proving 
\eqref{planning:condition} is equivalent to showing that for each $l\in\I_i$, 
$S_l^i\cap B(\chi_i(\delta t);r_i)\ne\emptyset$ iff there exists $w_i\in W_i$ 
so that $k_{i,\bf l_i}$, $w_i$, $l$ satisfy the Transition Requirement. For the 
given discretization, it follows from 
Proposition~\ref{prop:discretizations:keep:k:unsaturated} and standard ODE 
arguments that the solution $z_i(\cdot)$ of \eqref{system:disturbances} with  
$v_i(t)=k_{i,\bf{l}_i}(t,z_i(t),\bf{d}_j(t);x_{i0},w_i)$ and 
$z_i(0)=x_{i0}$ coincides with the solution $\bar z_i(\cdot)$ of 
\eqref{system:disturbances} with  
$v_i(t)=\bar k_{i,\bf{l}_i}(t,\bar z_i(t),\bf{d}_j(t);x_{i0},w_i)$ and 
$\bar z_i(0)=x_{i0}$, for any continuous disturbances satisfying 
\eqref{disturbance:bounds}. Thus, to verify \eqref{planning:condition} we will 
equivalently show that 
\begin{align}
S_l^i\cap B(\chi_i(\delta t);r_i) \ne\emptyset\iff & \nonumber \\
\exists\, w_i\in W_i\;\textup{s.t.}\;\bar k_{i,\bf l_i}, & w_i,l\;{\rm 
satisfy}\;{\rm 
(TR)} \label{TR:equivalence:intersect}
\end{align}
for any $l\in\I_i$. To prove \eqref{TR:equivalence:intersect}, assume that 
$\bar k_{i,\bf l_i}$, $w_i$, $l$ satisfy (TR), implying that for any continuous 
disturbances such that \eqref{disturbance:bounds} holds, the solution 
$z_i(\cdot)$ of \eqref{system:disturbances} with  
$v_i(t)=\bar k_{i,\bf{l}_i}(t,z_i(t),\bf{d}_j(t);x_{i0},w_i)$ and 
$z_i(0)=x_{i0}\in S_{l_i}^i$ satisfies $z_i(\delta t)\in S_l^i$. From 
Lemma~\ref{lemma:unsaturated:control:trajectory}(i), we get that $z_i(\delta 
t)=\chi_i(\delta t)+\lambda(i)w_i\delta t$. Thus, we get from \eqref{set:W} and 
\eqref{distance:r} that $z_i(\delta t)\in B(\chi_i(\delta t);r_i)$, and since  
$z_i(\delta t)\in S_l^i$, that $S_l^i\cap B(\chi_i(\delta t);r_i) 
\ne\emptyset$. Conversely, let $x\in S_l^i\cap B(\chi_i(\delta t);r_i) 
\ne\emptyset$. Then, by selecting $w_i$ according to \eqref{vector:wi}, 
it follows from Lemma~\ref{lemma:unsaturated:control:trajectory}(ii) 
that $w_i\in W_i$ and that the solution $z_i(\cdot)$ of 
\eqref{system:disturbances} with  
$v_i(t)=\bar k_{i,\bf{l}_i}(t,z_i(t),\bf{d}_j(t);x_{i0},w_i)$ and 
$z_i(0)=x_{i0}\in S_{l_i}^i$ satisfies $z_i(\delta t)=x\in S_l^i$ for 
all $x_{i0}\in S_{l_i}^i$. Thus, $\bar k_{i,\bf l_i}$, 
$w_i$, $l$ satisfy (TR) and the proof is complete.     
\end{pf}

\begin{remark}
From the acceptable values of $\delta t$ and $d_{\max}(i)$ in 
\eqref{deltat:interval}, \eqref{dmax:interval}, it is observed that 
the discretizations become finer for larger dynamics bounds/Lipschitz 
constants, and tighter input constraints. Further, if follows 
from~\eqref{distance:r} that for larger values of the parameters $\lambda(i)$ 
it is possible to increase the number of successor transitions from each cell 
and enhance the abstraction accuracy by considering finer discretizations.
\end{remark}

\section{Correctness and Exploitation} 
\label{sec:correctness:applicability}

Having established the existence of well-posed discretizations, 
we will show that the composition of the corresponding discrete agent 
models provides a correct representation of \eqref{single:integrator} 
over the horizon $[0,T]$ and clarify how the abstractions can be 
used for control synthesis.

\subsection{Correctness of the Abstraction}

As the first step to prove correctness, we provide a lemma for 
well-posed discretizations, which guarantees that if a cell 
configuration of all agents intersects their exact reachable set at 
some $t^*\in[0,T-\delta t]$, any 
successor configuration will also intersect this set at $t^*+\delta t$. In 
addition, by assigning to each agent a feedback law which enables its 
transition, the solution of the closed loop coupled 
system~\eqref{single:integrator} will drive all agents to their successor cells 
at $\delta t$ from the designated initial conditions of their transitions. The 
proof is given in the Appendix.

\begin{lemma} \label{lemma:onestep:consistency}
\textbf{(One Step Correctness)} Consider a well-posed 
discretization $\{\S_i\}_{i\in\N}-\delta t$ 
and a cell configuration $\bf{l}=(l_1,\ldots,l_N)$ with 
$S_{l_i}^i\subset\R_i([0,T-\delta t])$ for all $i\in\N$. Also, assume 
that there exist a time $t^*\in[0,T-\delta t]$ 
and an input $\mbf v\in\U$ such that each component of the solution of 
\eqref{single:integrator} satisfies $x_i(t^*,\bf{X}_0;\mbf v)\in S_{l_i}^i$.  
Then,  for any $l_i'\in\overline{{\rm Post}}_i({\rm pr}_i(\bf{l}))$, 
$i\in\N$, the following hold. (i) There exist $v_{\max}(i)-W_i$ feedback laws 
$k_{i,{\rm pr}_i(\bf l)}(\cdot)$ and parameters $w_i\in W_i$, such that the 
solution $\mbf{\xi}(\cdot)$ of the closed loop system \eqref{single:integrator} 
with $v_i=k_{i,{\rm pr}_i(\bf l)}$ and initial condition $\mbf{\xi}(0)=\bf 
x(t^*,\bf{X}_0;\mbf v)$ satisfies $\xi_i(\delta t)\in S_{l_i'}^i$ for all 
$i$. (ii) There exists $\mbf u\in\U$, with $\mbf u(t)=\mbf v(t)$ for all 
$t\in[0,t^*)$, such that the solution of \eqref{single:integrator} satisfies 
$x_i(t^*+\delta t,\bf{X}_0;\mbf u)\in S_{l_i'}^i$ for all~$i$.
\end{lemma}
%
%We will use the result of Lemma~\ref{lemma:onestep:consistency} in order to 
%prove that the discrete paths generated by the product of the agents' abstract 
%models can be captured by the $\delta t$-sampled continuous 
%system~\eqref{single:integrator} on $[0,T]$ (recall that $T=\ell\delta t$ for 
%certain $\ell\in\mathbb N$). 
We next provide the definition of the product transition system based on 
the synchronization of the agents' actions in their individual transition 
systems.
\begin{dfn}\label{product:TS}
\textbf{(Product Transition System)} \textit{(i)} Consider a well 
posed discretization $\{\S_i\}_{i\in\N}-\delta t$, and each agent's individual 
transition system $TS_i$ as given by Definition \ref{individual:ts}. The 
product transition system $TS_{\P}:=(Q,Q_0,{\rm Act},\longrightarrow)$ consists 
of:  State set $Q:=\bfI=\I_1\times\cdots\times \I_N$; Initial state set 
$Q_0:=Q_{01}\times\cdots\times Q_{0N}$, 
$Q_{0i}:=\{l_i\in\I_i:X_{i0}\in S_{l_i}^i\}$, $i\in\N$; Actions ${\rm 
Act}:=\{[w_1]\times\cdots\times[w_N]:[w_i]\in 2^{W_i},i\in\N\}$; 
Transition relation $\longrightarrow\subset Q\times Act\times Q$ defined as 
follows. For any $\bf{l}=(l_1,\ldots,l_N),\bf l'=(l_1',\ldots,l_N')\in 
Q$, and $[w]=[w_1]\times\cdots\times[w_N]\in {\rm Act}$, 
$\bf{l}\overset{[w]}{\longrightarrow}\bf{l}'$ iff 
$l_i\overset{({\rm pr}_i(\bf l),[w_i])}{\longrightarrow_i}l_i'$ for all 
$i\in\N$.  \textit{(ii)} A path of length $m\in\{0,\ldots\ell\}$ 
originating from $\bf{l}^0 $ in $TS_{\P}$, is a finite sequence of states 
$\bf{l}^0 \bf{l}^1\cdots\bf{l}^{m}$ such that  $\bf{l}^0\in Q_0$ and $ 
\bf{l}^{\kappa}\in\overline{{\rm  Post}}(\bf{l}^{\kappa-1}):=\cup_{[w]\in 
Act}{\rm Post}(\bf l^{\kappa-1};[w])$ for all $\kappa\in\{1,\ldots,m\}$ if 
$m\ne 0$.
\end{dfn}
%
%\color{blue}
We next show that for well-posed discretizations, any path of length 
$m\in\{0,\ldots,\ell\}$ originating from certain $\bf{l}^0\in Q_0$ in 
$TS_{\P}$  can be realized by a sampled trajectory of the continuous-time 
system \eqref{single:integrator} on $[0,m\delta t]$, initiated from 
$\bf{X}_{0}$. In addition, when the cell decomposition of each set 
$\R_i([0,T])$ is compliant with $\R_i([0,T-\delta t])$,  
there always exists an outgoing transition from the final state of each path 
with length less than $\ell$, which also implies the existence of such paths. 
\begin{prop} \label{posposition:path:correctness}
\textbf{(Correctness of Discrete Paths)} Assume that the 
discretization is well posed and that $\bf{l}^0 \bf{l}^1\cdots\bf{l}^{m}$ is a 
path of length $m\in\{0,\ldots,\ell\}$ originating from $\bf{l}^0$ in 
$TS_{\P}$. Then, there exists an input $\mbf v\in\mathcal{U}$ such that each 
component $x_i(\cdot,\bf{X}_0;\mbf v)$ of the solution of 
\eqref{single:integrator} satisfies $x_i(\kappa\delta t,\bf{X}_0;\mbf v)\in 
S^i_{l_i^{\kappa}}$ for all $\kappa\in\{0,\ldots,m\}$. 
\end{prop}
\begin{pf}
The proof is performed by induction on segments of the path, i.e., by showing 
that for each $\kappa\in\{0,\ldots,m\}$ there exists  $\mbf v\in\mathcal{U}$ 
such that $x_i(\FK\delta t,\bf{X}_0;\mbf v)\in S^i_{l_i^{\FK}}$ for all 
$\FK\in\{0,\ldots,\kappa\}$ and $i\in\N$. For $\kappa=0$, this follows directly 
from the fact that each $X_{i0}\in S_{l_i^0}^i$, since $\bf l^0\in Q_0$. Assume 
next that the induction hypothesis is valid for certain 
$\kappa\in\{0,\ldots,m-1\}$, implying that there exists $\mbf v\in\mathcal{U}$ 
with $x_i(\kappa\delta t,\bf{X}_0;\mbf v)\in S^i_{l_i^{\kappa}}$, for all 
$i\in\N$. In addition, since $\bf l^{\kappa+1}\in\overline{{\rm  
Post}}(\bf{l}^{\kappa})$, we have from Definitions~\ref{individual:ts} 
and~\ref{product:TS} that $S_{l_i^{\kappa}}^i\subset\R_i([0,T-\delta t])$ for 
all $i\in\N$. Thus, application of Lemma~\ref{lemma:onestep:consistency}(ii) 
with $t^*=\kappa\delta t\le T-\delta t$, $\bf l=\bf l^{\kappa}$, and 
$l_i'=l_i^{\kappa+1}$, $i\in\N$ yields existence of an input $\mbf u\in\U$ with 
$\mbf u(t)=\mbf v(t)$ for all $t\in[0,\kappa\delta t)$ and 
$x_i((\kappa+1)\delta t,\bf{X}_0;\mbf u)\in S_{l_i^{\kappa+1}}^i$ for all 
$i\in\N$. The latter together with the induction hypothesis for $\kappa$ 
establishes validity of the general induction step, i.e., that $x_i(\FK\delta 
t,\bf{X}_0;\mbf u)\in S_{l_i^{\FK}}^i$ for all $\FK\in\{0,\ldots,\kappa+1\}$ 
and $i\in\N$.      
\end{pf}
\begin{corollary}
\textbf{(Existence of Discrete Paths)} Assume that the 
discretization is well posed and that each cell decomposition 
$\S_i=\{S_l^i\}_{l\in\I_i}$ of $\R_i([0,T])$ is compliant with 
$\R_i([0,T-\delta t])$. Then, the following hold. (i) For any path $\bf{l}^0 
\bf{l}^1\cdots\bf{l}^{m}$ of length $m\in\{0,\ldots,\ell-1\}$ it holds that  
$\overline{{\rm Post}}(\bf l^m)\ne\emptyset$, implying that for any 
$\bf l^{m+1}\in\overline{{\rm  Post}}(\bf l^m)$, $\bf l^0 \bf 
l^1\cdots\bf l^m\bf l^{m+1}$ is a path of length $m+1$. (ii) For any 
$m\in\{0,\ldots,\ell\}$, there exists a path $\bf{l}^0 
\bf{l}^1\cdots\bf{l}^{m}$ of length $m$ in $TS_{\P}$.
\end{corollary}
\begin{pf}
For the proof, let  $\bf{l}^0 \bf{l}^1\cdots\bf{l}^{m}$ be a path of length 
$m\in\{0,\ldots,\ell-1\}$. Then, we get from 
Proposition~\ref{posposition:path:correctness} that there exists $\mbf 
v\in\mathcal{U}$ such that $x_i(m \delta t,\bf{X}_0;\mbf v)\in S^i_{l_i^m}$ for 
all $i\in\N$. Thus, since $m\delta t\le T-\delta t$, we get 
from~\eqref{reach:set:inclusion} applied with $t=T-\delta t$ that 
$S^i_{l_i^m}\cap\R_i([0,T-\delta t])\ne\emptyset$. Hence, by compliance (see 
end of Section~\ref{sec:preliminaries}), it follows that 
$S^i_{l_i^m}\subset\R_i([0,T-\delta t])$ for all $i\in\N$. Thus, since the 
discretization is well posed, we can select for each agent a cell 
$l_i^{m+1}\in\overline{{\rm  Post}}_i({\rm pr}_i(\bf l^m))$, which implies that 
$\bf l^{m+1}\in\overline{{\rm  Post}}(\bf l^m)$ and establishes validity of 
part (i). The proof of part (ii) follows directly from the inductive 
application of part (i) and the fact that $\bf l^0$ is a path of length 0 in 
$TS_{\P}$.  
\end{pf}
\begin{remark} \label{remark:planning} (i) From Lemma 
\ref{lemma:onestep:consistency}(i) and arguing analogously as in the above 
proofs, we also deduce the useful fact that for any path $\bf{l}^0 
\bf{l}^1\cdots\bf{l}^{m}$ of length $m\in\{0,\ldots,\ell\}$ in $TS_{\P}$, there 
exist sequences of feedback controllers $k_{i,{\rm pr}_i(\bf l^0)}\ldots 
k_{i,{\rm pr}_i(\bf l^{m-1})}$, $i\in\N$, which can be applied successively at 
the time intervals $[0,\delta t],\dots,[(m-1)\delta t,m\delta t]$ and guarantee 
that each agent $i$ will lie in cell $S_{l_i^{\kappa}}^i$ at $\kappa \delta t$, 
$\kappa=1,\ldots,m$. 
%Furthermore, the initial state set of each agent's transition system in 
%Definition~\ref{individual:ts} will be the indices of its decomposition's 
%cells 
%which intersect the corresponding set $\Omega_i$.
(ii) Consider an output map $H:\Rat{Nn}\to O$, with the set $O$ referring to 
properties of interest associated to the agents' joint state space, satisfying 
the compliance property that for any cell configuration $\bf 
l=(l_1,\ldots,l_N)$ and $\bf x,\bf x'\in\Rat{Nn}$ with 
$x_i,x_i'\in S_{l_i}^i$ for all $i\in\N$, it holds that 
$H(\bf x)=H(\bf x')$, and define $\bar H:\mbf{\mathcal I}\to O$ 
by $\bar H(\bf l)=H(\bf x)$ for any $\bf x$ with $x_i\in 
S_{l_i}^i$. Then, for any	output path $\mbf o^1 \mbf 
o^2\cdots \mbf o^m$ of length $m\le \ell$, i.e., 
such that $\mbf o^{\kappa}=\bar H(\mbf l^{\kappa})$ for each 
$\kappa$ and some path $\bf{l}^0 \bf{l}^1\cdots\bf{l}^{m}$ in 
$TS_{\P}$, there exist sequences of feedback controllers as 
in (i) above, so that the solution $\mbf{\xi}(\cdot,\bf X_0)$ 
of the closed loop system~(1) satisfies 
$H(\mbf{\xi}(\kappa\delta t,\bf X_0))=\mbf o^{\kappa}$ for 
each $\kappa$.
(iii) The analysis and results of Sections~\ref{sec:abstractions} and 
\ref{sec:correctness:applicability} are also applicable when each agent's 
initial states belong to a bounded set $\Omega_i$, with 
Assumption~\ref{assumption:completeness} guaranteeing again the existence of 
bounded overapproximations for the agents' reachable sets from  
$\mbf{\Omega}:=\Omega_1\times\cdots\times\Omega_{N}$ for the abstractions' 
derivation.
\end{remark}

\subsection{Exploiting the Abstractions for Control Synthesis} 
\label{sec:exploitation}

Consider a set of specifications assigned to the agents, whose 
satisfiability can be checked over a bounded time horizon $[0,T]$. Then, the 
abstractions from Theorem~\ref{dfn:well:posed:discretization} over this horizon 
can be exploited for control synthesis under these tasks through the following 
procedure.

\noindent \textit{Step 1.} Given the agents' initial positions and the time 
horizon, determine overapproximations of their reachable sets as in 
\eqref{overapproximations:Ri} 
and evaluate the corresponding bounds $M(i)$ and Lipschitz constants $L_1(i)$, $L_2(i)$ of the auxiliary dynamics. 

\noindent  \textit{Step 2.} Based on the evaluated bounds and Lipschitz 
constants from Step~1, and the input bounds $v_{\max}(i)$, select design 
parameters $\lambda(i)$, $\mu(i,j)$ and leverage 
Theorem~\ref{dfn:well:posed:discretization} to obtain a time step $\delta t$ 
and cell decompositions $\S_i$ of a well-posed discretization.

\noindent \textit{Step 3.} Fix a reference point for every cell of the 
decompositions and derive the transition system $TS_i$ of each agent $i$ as 
follows. For each cell configuration $\bf{l}_i$ compute the endpoint 
$\chi_i(\delta t)$ of the reference trajectory \eqref{reference:trajectory} and 
specify the cells which intersect $B(\chi_i(\delta t);r_i)$ to obtain the 
transitions to the cells given by the right-hand side of 
\eqref{planning:condition}. Note that the reference trajectories do not need to 
be stored and that no control laws are calculated at this step.

\noindent \textit{Step 4.} Find satisfying discrete plans whose composition 
generates a path $\bf{l}^0\bf{l}^1\bf{l}^2\cdots$ in the product transition 
system $TS_{\P}$ of Definition~\ref{product:TS}, that projects to a sequence of 
transitions 
$l_i^0\overset{(\bf{l}_i^0,[w_i]^0)}{\longrightarrow_i}l_i^1\overset{(\bf{l}_i^1,[w_i]^1)}{\longrightarrow_i}l_i^2\cdots$
 for each agent.

\noindent \textit{Step 5.} Select the continuous control laws to implement each 
agent's discrete plan as follows. For each transition 
$l_i^m\overset{(\bf{l}_i^m,[w_i]^m)}{\longrightarrow_i}l_i^{m+1}$, recalculate 
the trajectory $\chi_i(\cdot)$ corresponding to the cell configuration 
$\bf l_i^m$, pick any $x\in B(\chi_i(\delta t);r_i)\cap S_{l_i^{m+1}}^i$,  
compute $w_i$ from \eqref{vector:wi}, and select the corresponding 
feedback $k_{i,\bf l_i^m}(\cdot,\cdot,\cdot;\cdot,w_i)$ from 
\eqref{feedback:ki}.

Recalculating the reference trajectories and evaluating the control laws in 
Step 5 can significantly alleviate the memory storage requirements of the 
abstraction scheme, since these operations are performed only for the 
transitions of the satisfying plan and not for all the transitions 
of the abstractions. Examples of specifications whose satisfiability can 
be checked in finite time include metric interval temporal logic (MITL) 
formulas of the form $\Diamond_{[1,1.5]}(R_1\land\Diamond_{[2.5,3.5]}R_2)$, 
meaning ``visit region $R_1$ within 1 and 1.5 time units, and once $R_1$  has 
been visited, visit region $R_2$ within 2.5 and 3.5 time units", whose validity 
can be checked over the finite horizon $[0,5]$. Here, the timed eventually 
operator $\Diamond_{[a,b]}$ is used to denote that formula 
$\Diamond_{[a,b]}\phi$ is 
valid if $\phi$ becomes true at some time between $a$ and $b$ time units (see 
e.g.,~\cite{NaBdTjDd18} for more details on the MITL semantics).
Thus, desirable specifications include formulas with bounded temporal 
operators and any horizon of a length larger than the maximum over all sums of 
the upper bounds among nested temporal operators is sufficient to determine 
satisfiability of the formula (see \cite{RvDaMmMretal14}). 

\section{Example and Simulations} \label{sec:example}

\begin{figure}[tbh]
	\vspace{-.5em}
	\begin{center}
		\includegraphics[width =.8\columnwidth]{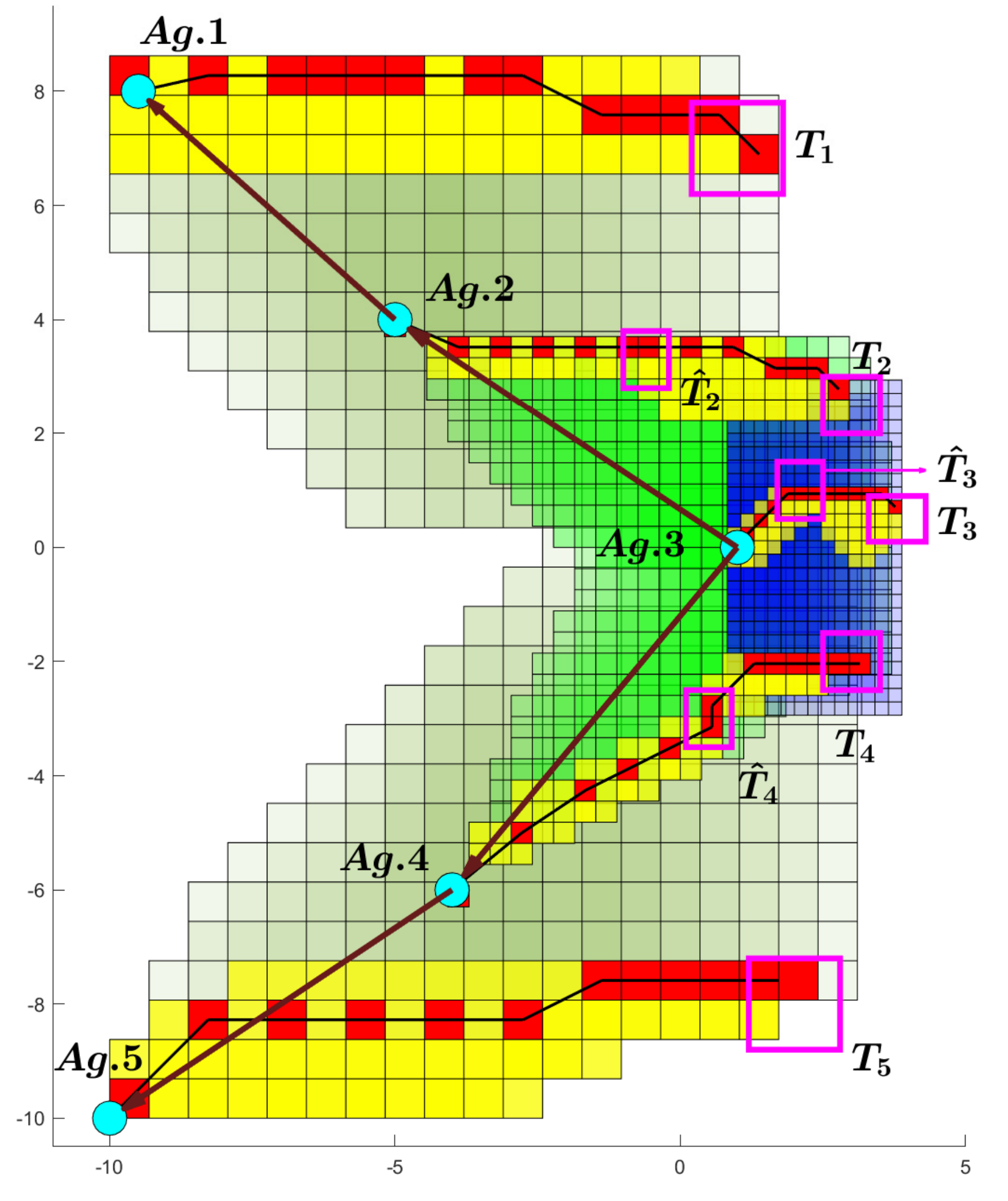} 
	\end{center}
\vspace{-1em}
\caption{The agents' initial positions are depicted with the cyan circles and 
the network topology is shown with the brown arrows. Each agent 
$i$ needs to reach the region(s) $T_i$ (and $\hat T_i$, $i=2,3,4$) depicted 
with the magenta box(es) within its specification deadlines. The selected 
discrete paths that satisfy all the agents' specifications are depicted in red. 
The cells of agent 3 that satisfy its task are shown in yellow and the 
reachable cells of agents 2 and 4 for the selected path of 3 are in 
green. The cells of agents 2 and 4  that satisfy their tasks are shown in 
yellow and all corresponding reachable cells of agents 1 and 5 are in light 
green, with those satisfying further their specifications depicted in yellow.} 
\label{figure:simulations}
\end{figure}

As an illustrative example we consider the coordination of five interconnected 
agents in $\Rat{2}$. Agent 3 represents an autonomous ground vehicle 
operating on an inclined terrain formed by a hill centered at zero,  whose 
height is given by $h(x):=C(1+\cos(\frac{\pi}{R}|x|))$, for $|x|<R$ and 
$h(x)=0$, for $|x|\ge R$. Its dynamics are $\dot x_3=f_3(x_3)+v_3$, with 
$f_3(x_3):=-\nabla h(x_3)$.
The other agents are UAVs 
operating at the same height and thus, modeled with two dimensional dynamics. 
Agents 2 and 4 are coupled with agent 3 by $\dot x_i=x_3-x_i+v_i$, $ 
i=2,4$, and agents 1 and 5 are coupled with 2 and 4 by $\dot x_i=x_j-x_i+v_i$, 
$(i,j)=(1,2),(5,4)$. The agents' additive inputs satisfy 
$v_{\max}(3)\le\frac{r}{2}$, $v_{\max}(i)\le r$, $i=2,4$,  and $v_{\max}(i)\le 
2r$, $i=1,5$, for certain $r\ge \frac{2C\pi}{R}$. The interconnection 
dynamics guarantee that agents 2, 3 and 3, 4, that are initially located at a
distance no more than $2r$, maintain this distance bound during the system's 
evolution. This allows them to exchange information for their feedback loop 
as well as high-level missions that are assigned during runtime, and can be 
verified by using the energy functions 
$V_{i3}(x_i,x_3)=\frac{1}{2}|x_i-x_3|^2$, $i=2,4$. Analogously, agents 1, 2 and 
4, 5 maintain a distance of at most $5r$, where we consider enhanced sensing 
capabilities among aerial agents compared to the ones among UAVs 
and the ground vehicle. Hence, the overall network, 
whose topology is depicted in Fig.~\ref{figure:simulations}, remains connected 
during the system's evolution for all input signals bounded by the given 
$v_{\max}(i)$. At some time instant (w.l.o.g. at $t=0$), the agents 
are assigned a set of timed specifications to reach the target boxes in 
Fig.~\ref{figure:simulations}. We assign the tasks $\Diamond_{[0.95,1]}(\hat 
T_i\land\Diamond_{[0.95,1]}T_i)$  to agents $i=2,4$ (reach $\hat T_i$ between 
0.95 and 1 time units, and once $\hat T_i$ is reached, reach $T_i$ between 0.95 
and 1 time units), $\Diamond_{[1.95,2]}T_i$ to agents $i=1,5$, and 
$\Diamond_{[0.85,1.05]}\hat T_3\land\Diamond_{[1.95,2]}T_3$ to agent 3 (reach 
$\hat T_3$ between 0.85 and 1.05 time units, and $T_3$ between 1.95 and 2 time 
units). Based on the discussion at the end of 
Section~\ref{sec:exploitation}, we select the horizon length $T=2$ that is 
sufficient to infer about satifiability of the tasks. From the bounds on the  
inter-agent distances, we obtain the following overapproximations of their 
reachable sets: $\R_{\bar{\N}_3}([0,T]) =\R_3([0,T]):=X_{30}+B(rT)$; 
$\R_{\bar{\N}_i}([0,T]):=\{(x_i,x_3)\in\Rat{4}:x_3\in\R_3([0,T])\;{\rm 
and}\;|x_i-x_3|<2r\}$, $\R_i([0,T]):=X_{i0}+B(3rT)$, $i=2,4$; $
\R_{\bar{\N}_i}([0,T]):=\{(x_i,x_j)\in\Rat{4}:x_j\in\R_j([0,T])\;{\rm 
and}\;|x_i-x_j|<5r\}$, $(i,j)=(1,2),(5,4)$, $R_i([0,T]):=X_{i0}+B(8rT)$, 
$i=1,5$. Using these sets and assuming that the parameters $C$ and $R$ in the 
dynamics of agent 3 satisfy $C\pi^2/R^2\le 1$, we obtain the bounds 
$M(3)=\frac{r}{2}$, $M(i)=2r$, $i=2,4$, $M(i)=5r$, $i=1,5$ and Lipschitz 
constants $L_1(3)=0$, $L_2(3)=3$, and $L_1(i)=L_2(i)=1$, $i=1,2,4,5$. 
For the simulations we select the parameters $r=5$, $C=2$, and $R=2\pi$, 
related to the agents dynamics, the planning parameters 
$\lambda(i)=0.4$, $i=2,3,4$, $\lambda(i)=0.35$, $i=1,5$, and $\mbf{\mu}(i)=1$ 
for all $i$, $\delta t=2/12$, i.e., $\ell=12$ time steps, and the largest 
corresponding values of $d_{\max}(i)$. 
%
%The agents' initial conditions are depicted with the cyan circles in 
%Fig.~\ref{figure:simulations} and satisfy their initial relative distance 
%requirement. 
%
%All agents need 
%to reach their target boxes $T_i$ depicted in Fig.~\ref{figure:simulations} at 
%$t=2$. Agents 2 and 4 are assigned also the specification to reach the boxes 
%$\hat T_2$ and $\hat T_4$, respectively, at $t=1$, and agent 3 to reach box 
%$\hat T_3$ at some time $t\in[0.85,1.05]$.     

First, we use the individual transition system of agent 3 to determine all the 
cells reachable by the agent over the horizon, shown in blue in 
Fig.~\ref{figure:simulations}, and then select all the paths 
which satisfy its reachability specifications, depicted through the 
corresponding yellow cells. We next select one of these paths, illustrated 
through the red cells of the agent in the figure, and exploit 
the individual transition systems of agents 2 and 4, in order to determine 
their reachable cells for the  selected path $l_3^0l_3^1\cdots l_3^{\ell}$ of 
agent 3. In particular, by denoting as $l_2^0$ the index of the cell where the 
initial state $X_{20}$ of agent 2 belongs, we evaluate the indices of its 
reachable cells at time $\kappa\delta 
t$ as $Q_2^{\kappa}={\rm Post}_2(Q_2^{\kappa-1},l_3^{\kappa-1})$, 
$\kappa=0,1,\ldots,\ell$. Note that $Q_2^0:=\{l_2^0\}$ and we have used the 
notation ${\rm Post}_2(Q_2,l_3):=\cup_{l_2\in Q_2}{\rm Post}_2(l_2;(l_2,l_3))$ 
(recall that $(l_2,l_3)$ stands for the cell configuration of agent 2).
Analogously, we evaluate the reachable cells of agent 4, depicted together with 
the ones of 2 with green in Fig.~\ref{figure:simulations}. Next, we determine 
the cells of 2 that reach its target $\hat T_2$ between 0.95 and 1 time 
units, and restrict its subsequent forward reachable cells to be originating 
from $\hat T_2$. After determining the ones from the latter that also 
reach the target $T_2$ within the next deadline, we perform a backward 
reachability algorithm and determine all the paths of agent 2 which satisfy 
both its reachability tasks, depicted with the corresponding yellow cells in 
Fig.~\ref{figure:simulations} (analogously for agent 4). In a similar 
manner we derive the transitions of agents 1 and 5 corresponding to all 
the satisfying trajectories of 2 and 4, respectively, depicted with the 
light green cells in the figure, and compute all the discrete paths of 
1 and 5 that satisfy their tasks, illustrated in yellow. Finally, we 
select a satisfying discrete trajectory for each of the agents 1, 2, 4, 
5, depicted through the red cells in Fig.~\ref{figure:simulations}. The 
simulation results have been implemented in MATLAB with a running time 
of 15.395486 seconds, on a PC with an Intel(R) Core(TM) i5-4210U @ 
1.7GHz processor. 
It is observed that the specifications can facilitate complexity 
reduction by  restricting the satisfying paths of parent agents that are used 
as actions to compute the associated part of their children's abstractions.

\section{Conclusions and Future Work} \label{sec:conclusions}

We introduced an online abstraction framework which guarantees the 
existence of symbolic models for forward complete multi-agent systems under 
coupled constraints. For each agent, we derive an 
individual discrete model for an overapproximation of its reachable set over a 
finite time horizon. Further, the composition of the individual 
abstractions captures the evolution of the continuous-time 
system over the horizon. Ongoing work directions include the consideration of 
graph structures which can reduce the synthesis complexity and the 
consideration of agents with higher order dynamics. It is additionally 
desirable to address robustness 
of the transitions~\cite{RgWaRm17} and integrate the 
abstraction framework with receding horizon planing under high-level 
specifications~\cite{TjDd16}.

\appendix
\section{Appendix}

Here we provide the proofs of 
Lemmas~\ref{lemma:overapproximations},~\ref{lemma:bounds}, 
and~\ref{lemma:onestep:consistency}, and 
Proposition~\ref{prop:discretizations:keep:k:unsaturated}. 
\begin{pf}\bf{(of Lemma~\ref{lemma:overapproximations})}
The proof of \eqref{reach:set:equality} follows directly from \eqref{overapproximations:Ri}, \eqref{cis:dfn},  and associativity of Minkowski addition. Indeed, we have that $\R_i([0,t])+B(c_i(T-t))=\R_i([0,T-\tau])+B(c_i(t-T+\tau))+B(c_i(T-t))=\R_i([0,T-\tau])+B(c_i(\tau))=\R_i([0,T])$.
In order to show~\eqref{reach:set:inclusion}, let $s\in[0,t]$ and $\mbf v\in 
\U$. Then, if $s\in[0,T-\tau]$, we have that $x_i(s,\bf X_0;\mbf v)\in 
\R_i([0,T-\tau])$ and thus, by \eqref{overapproximations:Ri}, also that  
$x_i(s,\bf X_0;\mbf v)\in \R_i([0,t])$. Next, let $s\in [T-\tau,t]$. Since 
$t\le T$, we obtain from \eqref{iandneighbors:inclusion} and 
\eqref{dynamics:bound} that $|\int_{T-\tau}^s(f_i(x_i(\sigma,\bf X_0;\mbf 
v),\bf x_j(\sigma,\bf X_0;\mbf 
v))+v_i(\sigma))d\sigma|\le(M(i)+v_{\max}(i))(s-T+\tau)$. Thus, since $s\le t$, 
it follows from \eqref{cis:dfn} that $\int_{T-\tau}^s(f_i(x_i(\sigma,\bf 
X_0;\mbf v),\bf x_j(\sigma,\bf X_0;\mbf v))+v_i(\sigma))d\sigma\in 
B(c_i(t-T+\tau))$. From the latter,  \eqref{overapproximations:Ri}, and 
the fact that $x_i(T-\tau,\bf X_0;\mbf v)\in\R_i([0,T-\tau])$ and $x_i(s,\bf 
X_0;\mbf v)=x_i(T-\tau,\bf X_0;\mbf v)+\int_{T-\tau}^s(f_i(x_i(\sigma,\bf 
X_0;\mbf v),$ $\bf x_j(\sigma,\bf X_0;\mbf v))+v_i(\sigma))d\sigma$,  we get 
that 
\eqref{reach:set:inclusion} is fulfilled and the proof is complete.
\end{pf}
\begin{pf} \bf{(of Lemma~\ref{lemma:bounds})}
By taking into account \eqref{function:g} and that the range of $k_{i,\bf 
l_i}$  lies in $B(v_{\max}(i))$, the dynamics of \eqref{system:disturbances} 
are bounded. Combined with their Lipschitz property, this implies that 
$z_i(\cdot)$ is well defined for all times and satisfies 
$|z_i(t)-x_{i0}|\le \int_0^t|g_i(z_i(s),\bf d_j(s))+k_{i,\bf 
l_i}(s,z_i(s),\bf d_j(s);x_{i0},w_i)|ds\le (M(i)+v_{\max}(i))t$. 
Consequently, we obtain from \eqref{cis:dfn} that $z_i(t)-x_{i0}\in 
B(c_i(\delta t))$ for all $t\in[0,\delta t]$. Thus, since $x_{i0}\in 
S_{l_i}^i\subset\R_i([0,T-\delta t])$, and due 
to~\eqref{reach:set:equality} it holds that $\R_i([0,T-\delta 
t])+B(c_i(\delta t))=\R_i([0,T])$, it follows that 
\begin{equation} \label{xi:in:RizeroT}
z_i(t)\in\R_i([0,T]), \forall t\in[0,\delta t].
\end{equation}
Analogously, since  $S_{l_{\kappa}}^{\kappa}\subset\R_{\kappa}([0,T-\delta 
t])$ for all $\kappa\in\N_i$, we obtain from~\eqref{disturbance:bounds} that
\begin{equation} \label{disturbances:in:RkappazeroT}
d_{\kappa}(t)\in\R_{\kappa}([0,T]), \forall t\in[0,\delta t],\kappa\in\N_i.
\end{equation}
In order to show \eqref{ki1:bound}, notice that due to \eqref{feedback:ki1} 
we have
\begin{align}
& 
k_{i,\bf{l}_i,1}(t,z_i(t),\bf{d}_j(t))=[g_i(\chi_i(t),\bf{x}_{l_j,G})-g_i(z_i(t),\bf{x}_{l_j,G})]
\nonumber \\
& \qquad +[g_i(z_i(t),\bf{x}_{l_j,G})-g_i(z_i(t),\bf{d}_j(t))].  
\label{ki1:equiv}
\end{align}
For the second difference on the right hand side of \eqref{ki1:equiv}, we 
obtain from \eqref{dynamics:bound1}, \eqref{xi:in:RizeroT}, and 
\eqref{disturbances:in:RkappazeroT} that  $|g_i(z_i(t),\bf{x}_{l_j,G}) - 
g_i(z_i(t),\bf{d}_j(t))| \le 
L_{1}(i)|(d_{j_{1}}(t)-x_{l_{j_1},G},\ldots,d_{j_{N_i}}(t)-x_{l_{j_N},G})|$.
From the latter combined with \eqref{cis:dfn},  \eqref{reference:points}, 
and \eqref{disturbance:bounds}, we get that $|g_i(z_i(t),\bf{x}_{l_j,G}) 
- g_i(z_i(t),\bf{d}_j(t))| \le 
L_{1}(i)(\sum_{\kappa\in\N_i}(\frac{d_{\max}(\kappa)}{2} 
+(M(\kappa)+v_{\max}(\kappa))t)^{2})^{\frac{1}{2}}$.
Thus, it follows from \eqref{diameters:restrictions}, \eqref{constant:mui}, 
\eqref{constant:bfM(i)}, and the Cauchy Schwartz inequality that 
\begin{align*}
& |g_i(z_i(t),\bf{x}_{l_j,G}) -  g_i(z_i(t),\bf{d}_j(t))| \\
& \le   L_{1}(i)[\sum_{\kappa\in\N_i}(\mu(\kappa,i)d_{\max}(i)/2
+(M(\kappa)+v_{\max}(\kappa))t)^{2}]^{\frac{1}{2}}  \\
& \le 
L_{1}(i)[d_{\max}(i)^2/4\sum_{\kappa\in\N_i}\mu(\kappa,i)^2+d_{\max}(i)t(\sum_{\kappa\in\N_i}\mu(\kappa,i)^2)^{\frac{1}{2}}\\
& \times (\sum_{\kappa\in\N_i}(M(\kappa)+v_{\max}(\kappa))^2)^{\frac{1}{2}}+t^2 
\sum_{\kappa\in\N_i}(M(\kappa)+v_{\max}(\kappa))^2]^{\frac{1}{2}} \\
%\end{align*}
%
%\begin{align*}
& = 
L_{1}(i)[(\sum_{\kappa\in\N_i}\allowbreak\mu(\kappa,i)^2)^{\frac{1}{2}}d_{\max}(i)/2
+(\sum_{\kappa\in\N_i}(M(\kappa) \\
& +v_{\max}(\kappa))^2)^{\frac{1}{2}}t]= 
L_{1}(i)(\mbf{\mu}(i)d_{\max}(i)/2+\bf{M}(i)t)
\end{align*}
for all $t\in[0,\delta t]$. For the other difference in \eqref{ki1:equiv}, 
it follows analogously to the derivation of~\eqref{xi:in:RizeroT} that 
$\chi_i(t)\in\R_i([0,T])$ for all $t\in[0,\delta t]$. From the latter 
combined 
with~\eqref{xi:in:RizeroT} and \eqref{dynamics:bound2} we get that 
$|g_i(z_i(t),\bf{x}_{l_j,G})-g_i(\chi_i(t),\bf{x}_{l_j,G})|\le 
L_2(i)|z_i(t)-\chi_i(t)|$ for all $t\in[0,\delta t]$. Hence, it follows 
from 
the evaluated bounds on the differences in \eqref{ki1:equiv} that 
\eqref{ki1:bound} holds. Next, from \eqref{feedback:ki2} and 
\eqref{set:W} we get that \eqref{ki2:bound} holds. Finally, by recalling that 
$x_{l_i,G}$ satisfies \eqref{reference:points}, it follows from 
\eqref{feedback:ki3} that \eqref{ki3:bound} is fulfilled.
\end{pf}
\begin{pf}
\textbf{(of Proposition~\ref{prop:discretizations:keep:k:unsaturated})} 
To prove the proposition, let $\bar t:= \sup\{t\in[0,\delta t]:\forall 
s\in[0,t],|\bar k_{i,\bf l_i}(s,z_i(s),$ $\bf 
d_j(s);x_{i0},w_i)|\le v_{\max}(i)\}$ and note that due to \eqref{feedback:ki}, 
it suffices to show that $\bar t=\delta t$. By 
recalling that   $S_{l_{\kappa}}^{\kappa}\subset\R_{\kappa}([0,T-\delta 
t])$ for all $\kappa\in\N_i$ and that $d_{j_1},\ldots,d_{j_{N_i}}$ 
satisfy  \eqref{disturbance:bounds}, it follows from 
Lemma~\ref{lemma:bounds} that the components $k_{i,\bf{l}_i,1}(\cdot)$, 
$k_{i,\bf{l}_i,2}(\cdot)$, and $k_{i,\bf{l}_i,3}(\cdot)$ 
in~\eqref{feedback:ki} evaluated along the solution 
of~\eqref{system:disturbances} satisfy the bounds 
\eqref{ki1:bound}--\eqref{ki3:bound}. Thus, due to 
\eqref{ki1:bound}, \eqref{reference:trajectory}, and \eqref{reference:points}, 
we have that $|k_{i,\bf{l}_i,1}(0,z_i(0),\bf{d}_j(0))|\le 
\frac{(L_{1}(i)\mbf{\mu}(i)+L_2(i))d_{\max}(i)}{2}$. The latter together 
with \eqref{ki3:bound} and \eqref{ki2:bound} imply by 
virtue of \eqref{dmax:interval} that $|\bar 
k_{i,\bf{l}_i}(0,z_i(0),\bf{d}_j(0);\allowbreak 
x_{i0},w_i)|<v_{\max}(i)$. Thus, $[0,\bar t]$ is well defined. We next show 
that
\begin{equation} \label{controller:strict:ineq:bound}
|\bar 
k_{i,\bf{l}_i}(t,z_i(t),\bf{d}_j(t);x_{i0},w_i)|<v_{\max}(i),\forall 
t\in[0,\bar t],
\end{equation}
which implies that $\bar t=\delta t$. Indeed, if $\bar t<\delta t$, we 
would have by \eqref{controller:strict:ineq:bound} and continuity of 
$t\mapsto \bar k_{i,\bf{l}_i}(t,z_i(t),\bf{d}_j(t);\allowbreak 
x_{i0},w_i)$ that $|\bar 
k_{i,\bf{l}_i}(t,z_i(t),\bf{d}_j(t);x_{i0},w_i)|<v_{\max}(i)$ for all 
$t\in[0,\bar t+\varepsilon)$ and certain $\varepsilon\in (0,\delta t-\bar 
t)$, contradicting the definition of $\bar t$. In order to show 
\eqref{controller:strict:ineq:bound}, note that due to the definition of $\bar 
t$, it holds for all $t\in[0,\bar t]$ that 
$k_{i,\bf{l}_i}(t,z_i(t),\bf{d}_j(t);x_{i0},w_i)=\bar 
k_{i,\bf{l}_i}(t,z_i(t),\bf{d}_j(t);x_{i0},\allowbreak w_i)$. From the 
latter and Lemma~\ref{lemma:unsaturated:control:trajectory} we get 
that \eqref{xi:trajectory} holds. Combining this with 
\eqref{reference:points}, \eqref{set:W}, and 
\eqref{ki1:bound}, we obtain that 
$|k_{i,\bf{l}_i,1}(t,z_i(t),$ $\bf{d}_j(t))| \le 
L_{1}(i)(\mbf{\mu}(i)\frac{d_{\max}(i)}{2}+\bf{M}(i)t) 
+L_{2}(i)(\frac{(\delta t-t)d_{\max}(i)}{2\delta 
t}\allowbreak+\lambda(i)v_{\max}(i)t)$, for all $t\in[0,\bar t]$. By taking 
into account the latter, \eqref{ki2:bound}, and 
\eqref{ki3:bound}, in  
order to show \eqref{controller:strict:ineq:bound}, we need to prove that 
$L_{1}(i)(\mbf{\mu}(i)\frac{d_{\max}(i)}{2}+\bf{M}(i)t)+\frac{d_{\max}(i)}{2\delta
t}+L_{2}(i)(\frac{(\delta t-t)d_{\max}(i)}{2\delta 
t}+\lambda(i)v_{\max}(i)t)+\lambda(i) v_{\max}(i)< v_{\max}(i)$, for all 
$t\in[0,\bar t]$. Since the latter is linear with respect to $t$ and 
$[0,\bar t]\subset [0,\delta t]$, it suffices to verify it for $t=0$ and 
$t=\delta t$. Both cases follow directly from \eqref{dmax:interval} and 
the proof is complete.
\end{pf}
\begin{pf}\bf{(of Lemma \ref{lemma:onestep:consistency})}
For the proof of both assertions let $l_i'\in\overline{{\rm Post}}_i({\rm 
pr}_i(\bf l))$ for each agent. Hence, given the cell configuration ${\rm 
pr}_i(\bf{l})$ of each $i$, there exist a $v_{\max}(i)-W_i$ feedback law 
$v_i=k_{i,{\rm pr}_i(\bf l)}(t,x_{i},\bf{x}_j;x_{i0},w_i)$, and a vector 
$w_i\in W_i$, such that $k_{i,{\rm pr}_i(\bf l)}$, $w_i$, $l_i'$ satisfy the 
Transition Requirement (recall that the later refers to the system with 
disturbances \eqref{system:disturbances}). Next, we select for each agent the 
initial condition $x_{i0}=x_i(t^*,\bf{X}_0;\mbf{v})$ and denote by 
$\mbf{\xi}(\cdot)$ the solution of the closed loop system 
\eqref{single:integrator} with $v_i=k_{i,{\rm pr}_i(\bf{l})}$, $i\in\N$ as 
selected above. By the Lipschitz properties on the $f_i$'s and $k_{i,{\rm 
pr}_i(\bf{l})}$'s, this solution is unique and defined on 
the right maximal interval $[0,T_{\max})$. We will show that $T_{\max}=\infty$, 
and that for each agent $i\in\N$, the $i$-th component $\xi_i(\cdot)$ of 
$\mbf{\xi}(\cdot)$ coincides with the solution of system 
\eqref{system:disturbances} on $[0,\delta t]$ with 
$\bf{d}_j(t)=\mbf{\xi}_j(t)(=(\xi_{j_1}(t),\ldots,\xi_{j_{N_i}}(t))$ 
and the same initial condition $x_{i0}$. Indeed, consider the input 
$\mbf{\omega}=(\omega_1,\ldots,\omega_N):\RgeO\to\Rat{Nn}$ given as 
$\omega_i(t):=k_{i,{\rm 
pr}_i(\bf{l})}(t,\xi_{i}(t),\mbf{\xi}_j(t);x_{i0},w_i)$, $t\in[0,T_{\max})$ and 
$\omega_i(t)=0$, $t\ge T_{\max}$, for all $i\in\N$. Then, since the image 
of each $k_{i,\bf l_i}(\cdot)$ is contained in $B(v_{\max}(i))$, it follows 
that $|\omega_i(t)|\le v_{\max}(i)$ for all $t\ge 0$, and hence, that 
$\mbf{\omega}\in\U$. Thus, by denoting as $\mbf{\eta}(\cdot,\bf 
x_0;\mbf{\omega})$ the solution of \eqref{single:integrator} with input 
$\mbf{\omega}(\cdot)$ and initial condition $\bf x_0:=(x_{10},\ldots,x_{N0})$, 
we deduce from  Assumption~\ref{assumption:completeness} that it is defined for 
all positive times. In addition, it follows from standard ODE arguments that
\begin{equation}  \label{xi:eq:eta}
\mbf{\xi}(t)=\mbf{\eta}(t,\bf x_0;\mbf{\omega}),\forall t\in [0,T_{\max}).
\end{equation}
Thus, we get that $T_{\max}=\infty$, because otherwise it would hold that $\lim_{t\nearrow T_{\max}}\mbf{\xi}(t)=\mbf{\eta}(T_{\max},\bf{x}_0;\mbf{\omega})$, contradicting maximality of $[0,T_{\max})$. Next, define $\mbf u=(u_1,\ldots,u_N):\RgeO\to\Rat{Nn}$ as
\begin{align}
u_i(t):= & v_i(t), t\in[0,t^*), \nonumber \\
u_i(t):= & \omega_i(t-t^*), t\ge t^*  \label{input:ui}
\end{align}
\noindent for all $i\in\N$ and notice that $\mbf u\in\U$. Using standard 
properties of time-invariant deterministic control systems\footnote{see 
\cite[Chapter 2]{Se98}}, we obtain that $\mbf{\eta}(t,x_0;\mbf{\omega})=\bf 
x(t^*+t,\bf{X}_0;\mbf u)$, $\forall t\ge 0$. Hence, we get from  
\eqref{xi:eq:eta} that
\begin{equation} \label{xi:eq:ex}
\mbf{\xi}(t)=\bf x(t^*+t,\bf{X}_0;\mbf u), \forall t\ge 0
\end{equation}
%
%\color{blue} 
and since $t^*\le T-\delta t$, we obtain from \eqref{xi:eq:ex} and \eqref{iandneighbors:inclusion} that  
\begin{equation} \label{xi:property1}
(\xi_i(t),\mbf{\xi}_j(t))\in\R_{\bar{\N}_i}([0,T]) , \forall t\in[0,\delta t]
\end{equation}
for all $i\in\N$. Thus, we get from \eqref{xi:property1} and 
\eqref{dynamics:bound} that $|f_i(\xi_i(t),\mbf{\xi}_j(t))|\le M(i)$ for all 
$t\in[0,\delta t]$. The latter together with \eqref{cis:dfn} and the fact 
that $|k_{i,\bf l_i}(\cdot)|$ is bounded by $v_{\max}(i)$, implies that
\begin{equation} \label{xi:property3}
\xi_i(t)\in S_{l_i}^i+B(c_i(t)), \forall t\in[0,\delta t]
\end{equation}
for all $i\in\N$. In addition, we have from \eqref{xi:property1} and \eqref{gi:extension} that $g_i(\xi_i(t),\mbf{\xi}_j(t))=f_i(\xi_i(t),\mbf{\xi}_j(t))$ for all $t\in[0,\delta t]$, $i\in\N$. 
Thus, for each $i\in\N$, $\xi_i(\cdot)$ coincides on $[0,\delta t]$ with the 
solution of system \eqref{system:disturbances} with disturbances 
$\bf{d}_j(t)=\mbf{\xi}_j(t)$, input $v_i(t)=k_{i,\bf l_i}(t,\xi_i(t),\bf 
d_j(t);x_{i0},w_i)$, and the selected $x_{i0}$, $w_i$ at the beginning of the 
proof. Due to  \eqref{xi:property3}, the disturbances satisfy 
\eqref{disturbance:bounds}. Hence, by recalling that  
$S_{l_i}^i\subset\R_i([0,T-\delta t])$ and that the selected $k_{i,{\rm 
pr}_i(\bf l)}$, $w_i$, $l_i'$ satisfy the Transition Requirement, we deduce 
from Definition~\ref{conditionCC} that
%Thus, from the latter, \eqref{xi:property3}, the fact that $S_{l_i}^i\subset\R_i([0,T-\delta t])$ for all $i\in\N$, and the Transition Requirement applied to the system with disturbances \eqref{system:disturbances} with $\bf{d}_j=\mbf{\xi}_j$, initial condition $x_{i0}$, and the selected $w_i$ at the beginning of the proof, we deduce that 
\begin{equation}\label{xiatdeltat:inSli}
\xi_i(\delta t)\in S_{l_i'}^i
\end{equation}
for all $i\in\N$, which establishes part (i) of the lemma. Part (ii) 
follows directly by selecting $\mbf u(\cdot)$ as in \eqref{input:ui}, which by 
virtue of \eqref{xiatdeltat:inSli} and \eqref{xi:eq:ex} implies that 
$x_i(t^*+\delta t,\bf{X}_0;\mbf u)\in S_{l_i'}^i$ for all $i\in\N$. The proof 
is complete.
\end{pf}

\section{Acknowledgements}

This work was supported by the H2020 ERC Starting Grant BUCOPHSYS, the EU H2020 Co4Robots project, the Knut and Alice Wallenberg Foundation, the Swedish Foundation for Strategic Research (SSF), and the Swedish Research Council (VR). 

\bibliographystyle{unsrt}
\bibliography{longtitles,online_references}

\end{document}